\documentclass[10pt,letterpaper]{article}
\usepackage[top=0.85in,left=2.75in,footskip=0.75in]{geometry}

\usepackage{amsmath,amssymb}
\usepackage{mathtools} 

\usepackage{changepage}

\usepackage{textcomp,marvosym}

\usepackage{cite}

\usepackage{nameref,hyperref}

\usepackage[nopatch=eqnum]{microtype}

\usepackage{array}

\newcolumntype{+}{!{\vrule width 2pt}}

\newlength\savedwidth

\raggedright
\usepackage[aboveskip=1pt,labelfont=bf,labelsep=period,justification=raggedright,singlelinecheck=off]{caption}

\makeatletter
\renewcommand{\@biblabel}[1]{\quad#1.}
\makeatother

\usepackage{lastpage,fancyhdr,graphicx}
\usepackage{epstopdf}
\fancyheadoffset[L]{2.25in}
\fancyfootoffset[L]{2.25in}
\usepackage{booktabs}
\usepackage{multirow}
\usepackage{makecell}
\usepackage{float} 
\usepackage{algorithm}
\usepackage{algorithmic}
\usepackage{enumitem}
\usepackage{subcaption}
\usepackage{tabularx}
\usepackage{xcolor}  

\def\bX{\ensuremath{{\mathbf{X}}}}

\def\by{\ensuremath{{\mathbf{y}}}}
\def\bu{\ensuremath{{\mathbf{u}}}}

\def\bI{\ensuremath{{\mathbf{I}}}}

\def\bv{\ensuremath{{\mathbf{v}}}}

\usepackage[table]{xcolor}

\colorlet{highlightblue}{blue}   

\newcommand{\rev}[1]{{\color{highlightblue}#1}}

\colorlet{highlightblue}{blue}   

\newenvironment{revtableblock}
{%
  \begingroup
  \color{highlightblue}%
  \arrayrulecolor{highlightblue}%
}
{%
  \arrayrulecolor{black}%
  \endgroup
}

\colorlet{highlightred}{red}   

\newcommand{\revtwo}[1]{{\color{highlightred}#1}}

\newenvironment{revtableblocktwo}
{%
  \begingroup
  \color{highlightred}%
  \arrayrulecolor{highlightred}%
}
{%
  \arrayrulecolor{black}%
  \endgroup
}

\begin{document}
\vspace*{0.2in}

\begin{flushleft}
{\Large
\textbf{Data-Driven Modeling of Spatiotemporal Dynamics Using Multimodal Imaging Data}
}
\newline
\\
Chunyan Li\textsuperscript{1},
Yutong Mao\textsuperscript{2},
Xiao Liu\textsuperscript{2,3},
Wenrui Hao\textsuperscript{1*}
\\
\bigskip
\textbf{1} Department of Mathematics, The Pennsylvania State University, University Park, PA, USA
\\
\textbf{2} Department of Biomedical Engineering, The Pennsylvania State University, University Park, PA, USA
\\
\textbf{3} Institute for Computational and Data Sciences, The Pennsylvania State University, University Park, PA, USA
\\
\bigskip

* wxh64@psu.edu

\end{flushleft}

\section*{Abstract}
\begin{abstract}
Understanding how biological systems evolve across space and time remains a fundamental challenge, particularly when dynamic processes vary substantially across individuals. We present a personalized graph-based dynamical modeling framework for characterizing spatiotemporal biological dynamics from longitudinal multimodal imaging data. The framework constructs individualized brain graphs from MRI and PET measurements and learns patient-specific dynamical parameters governing regional structural and molecular changes. Applied to 1,891 participants from the Alzheimer’s Disease Neuroimaging Initiative, the model captures the coordinated evolution of amyloid-$\beta$, tau, neurodegeneration, and cognition and accurately predicts their future trajectories, outperforming established clinical and neuroimaging benchmarks. Patient-specific dynamical parameters reveal distinct patterns of biological progression and provide improved prediction of future cognitive decline compared with standard biomarkers. Sensitivity analysis further identifies regional network features associated with the propagation of pathological and structural changes, recovering known temporolimbic and frontal vulnerability patterns. These results demonstrate how data-driven dynamical modeling can integrate multimodal longitudinal measurements to uncover individualized spatiotemporal patterns and latent mechanisms of biological change. The framework provides a quantitative approach for studying complex biological dynamics across heterogeneous individuals and establishes a foundation for personalized modeling of progressive biological processes.\end{abstract}

\noindent\textbf{Keywords:} Alzheimer’s disease, Mathematical modeling, Neuroimaging, Sensitivity analysis



\section{Introduction}
\label{sec:introduction}
Alzheimer's disease (AD) affects over 50 million people worldwide and represents the most common form of dementia, imposing a profound societal, economic, and personal burden \cite{scheltens2021alzheimer}. Recent years have witnessed a landmark shift in the treatment landscape with the regulatory approval of disease-modifying therapies, specifically the anti-amyloid monoclonal antibodies lecanemab (Leqembi) and donanemab (Kisunla), which have been shown to slow cognitive decline in early-stage AD \cite{vanDyck2023lecanemab, sims2023donanemab, cummings2023pipeline}. However, these treatments only modestly slow progression rather than halt or reverse the disease, and their use is associated with challenges including amyloid-related imaging abnormalities (ARIA), high costs, limited accessibility, and the requirement for early diagnosis \cite{honig2023aria, cummings2023pipeline}. Thus, the need to better understand the underlying mechanisms driving AD onset and progression remains urgent. AD is characterized by extracellular amyloid-beta (A$\beta$) plaques and intracellular neurofibrillary tangles composed of hyperphosphorylated tau, which disrupt synaptic function, promote neuronal loss, and manifest as progressive cognitive decline \cite{hardy1992alzheimer, arezoumandan2022regional}. \rev{While the amyloid cascade hypothesis and related biomarker cascade frameworks have provided a foundational understanding of AD progression and are supported by substantial longitudinal imaging and clinical evidence \cite{selkoe2016amyloid, tolar2020path}, important questions remain regarding how these pathological processes propagate across interconnected brain regions and give rise to heterogeneous patient-specific trajectories. Addressing these challenges requires mechanistic models that integrate biomarker interactions with spatiotemporal disease dynamics.}

Recent advances in biomarker imaging have transformed our ability to study AD in vivo. Positron emission tomography (PET) allows visualization of A$\beta$ and tau deposition, structural MRI quantifies cortical atrophy, and cerebrospinal fluid (CSF) assays provide complementary molecular insights \cite{de2004mri, raj2024spectral, jack2017defining, scheltens2021alzheimer}. Functional imaging further reveals network-level alterations linked to cognitive impairment \cite{wiesman2021spatio}, emphasizing that the critical questions are not only \emph{what} changes occur in AD, but \emph{how}, \emph{where}, and \emph{why} pathology propagates across the brain. These data create an unprecedented opportunity to study AD as a spatiotemporally dynamic disease, but also highlight the limitations of conventional analytical approaches.

Mathematical and computational modeling \cite{shahriyari2022pde, kirshtein2020colon, le2021osteosarcoma, cang2018representability, kostelich2025mathematical, tursynkozha2025go, baez2016mathematical, huo2025oscillations} has emerged as an essential tool to integrate multi-modal data, formalize mechanistic hypotheses, and generate testable predictions and therapeutic interventions \textit{in silico} \cite{hao2016mathematical, bertsch2021amyloid, rabiei2025data, thompson2024alzheimer, cottrell2025computational, hao2022optimal}. Ordinary differential equation (ODE) models have been used to describe A$\beta$ production, aggregation, and clearance \cite{hao2016mathematical, bertsch2021amyloid, bossa2023multidimensional}, as well as tau hyperphosphorylation \cite{vosoughi2020mathematical, bertsch2023role}. Extensions incorporating neuroinflammation and other modulatory factors have provided additional mechanistic insights \cite{patel2024mathematical}, and partial differential equation (PDE) frameworks capture the spatial propagation of pathology across brain regions \cite{xu2025multiscale, hao2025optimal, raj2025understanding}. \rev{Complementing these mechanistic approaches, statistical and machine learning, and deep learning approaches have become increasingly important for analyzing large-scale clinical datasets. Mixed-effects models are widely used to characterize longitudinal trajectories and population heterogeneity, while machine learning and deep learning approaches leverage multimodal biomarkers to predict disease onset, stratify patients, and infer biomarker relationships \cite{zheng2022data, zhang2024discovering, wang2025learning, petrella2024personalized, davodabadi2023mathematical, petrella2019computational}. Although these methods often achieve strong predictive performance, they generally do not explicitly represent disease propagation over brain networks or provide a framework for evaluating competing biological hypotheses. As a result, they are generally less suited for testing mechanistic hypotheses or simulating intervention effects across interconnected brain regions within a unified disease-progression framework.}

Despite these advances, critical gaps remain. Most models focus on individual pathways or global biomarkers, failing to capture the integrated spatiotemporal dynamics of amyloid, tau, and neuroinflammation \cite{sandell2025integrating, sandell2025back, tora2025network, butler2023choroid, torok2025directionality}. Few frameworks provide personalized, spatially resolved predictions from longitudinal multi-modal imaging, limiting their translational relevance. Existing approaches often rely on MRI, regionally aggregated CSF or PET measures, lacking the spatial granularity needed to identify critical regions for targeted intervention \cite{vogel2020characterizing, sanami2025longitudinal}.
Here, we present a novel, mechanistic, data-driven framework to model the spatiotemporal progression of AD biomarkers at the individual level. Our approach formulates a system of PDEs on the brain’s functional connectivity network, capturing the regional propagation of three key AD biomarkers: A$\beta$, tau, cortical atrophy. The model incorporates subject-specific parameters inferred from PET and structural MRI data, enabling personalized predictions while preserving mechanistic interpretability. Comprehensive parameter inference and two-level sensitivity analyses identify critical disease drivers and brain regions. To our knowledge, this is the first mechanistic spatiotemporal model integrating multiple clinically validated biomarkers in a patient-specific framework. \rev{Beyond individualized prediction, the proposed framework serves as a mechanistic modeling platform that enables the incorporation and testing of competing biological hypotheses regarding disease propagation and therapeutic intervention effects.}

This framework provides three key advances:  
\begin{enumerate}
    \item It unifies temporal biomarker progression with spatial propagation across brain networks.  
    \item It enables patient-specific modeling of AD progression using multi-modal imaging biomarkers.  
    \item It facilitates the prediction of region-specific therapeutic responses, advancing precision medicine in AD.  
\end{enumerate}

By bridging mechanistic modeling, graph-based brain networks, and longitudinal neuroimaging, our approach provides a comprehensive, personalized view of AD progression, with direct implications for early diagnosis, individualized prognosis, and targeted therapeutic strategies.

\section{Results}
\label{sec:results}
\begin{figure}[!t]
    \centering
    \includegraphics[width=\columnwidth]{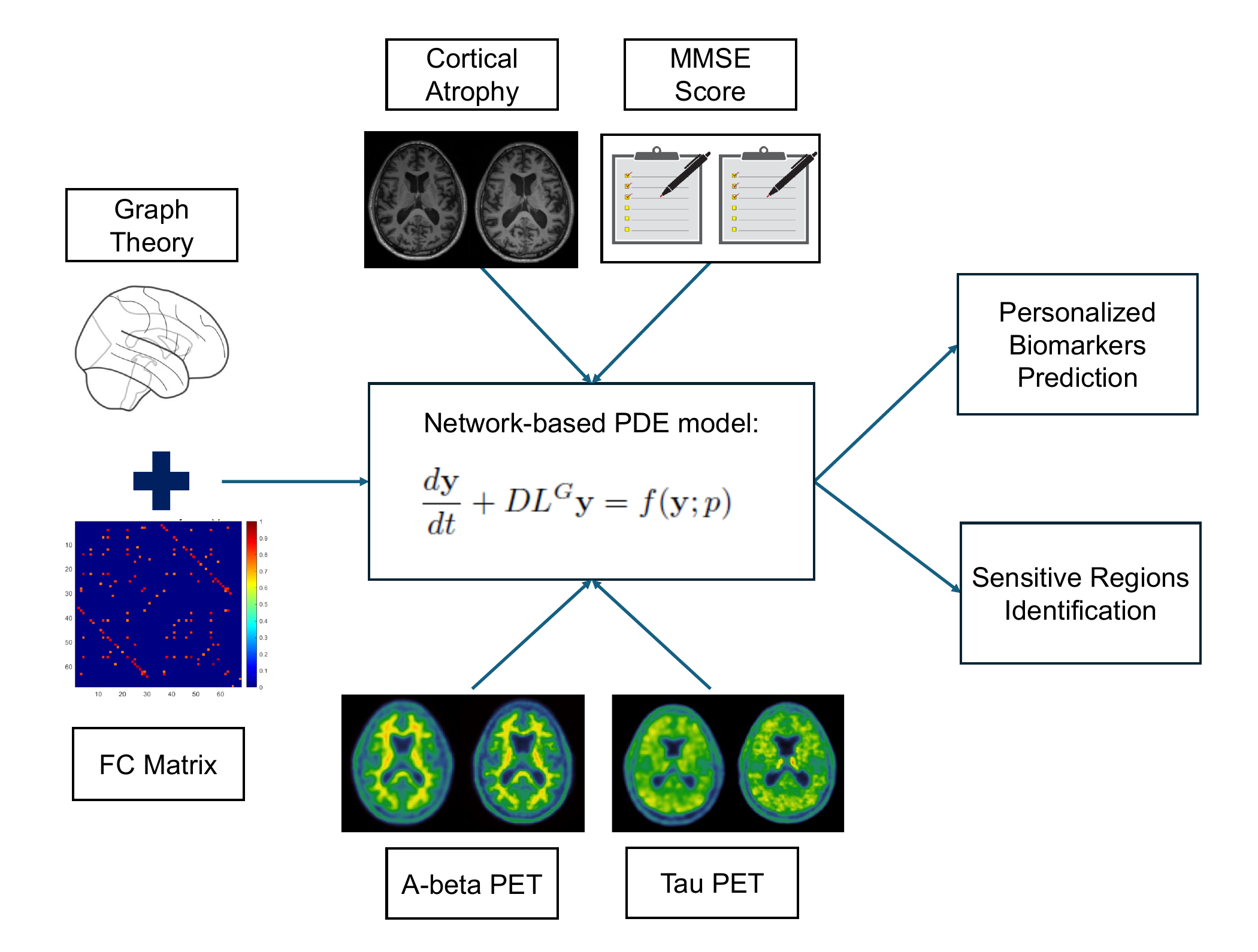}
     \caption{Workflow integrating graph theory, neuroimaging biomarkers (A$\beta$-PET, tau-PET, cortical atrophy, and MMSE score), AD pathophysiology, and mathematical modeling to construct a personalized digital twin of AD progression. The model, formulated as a spatiotemporal system of PDEs, captures biomarker diffusion and nonlinear interactions on the brain’s functional connectivity network. Coupled with sensitivity analysis, the framework enables personalized forecasting and identification of region-specific vulnerabilities with sensitivity analysis.}
    \label{fig:workflow}
\end{figure}

We applied the spatiotemporal generalized AD Biomarker Cascade (generalized ADBC) model to characterize individualized dynamics of clinically validated AD biomarkers and evaluate its potential for personalized disease forecasting. The model, governed by 14 parameters (6 global and 8 region-specific), captures heterogeneity across the brain’s functional network. Once estimated from neuroimaging data, it allows simulation of amyloid-beta (A$\beta$), tau, neurodegeneration (N), and cognitive decline (C), while a two-level sensitivity analysis identifies region-specific vulnerabilities. The overall workflow is illustrated in Fig.~\ref{fig:workflow}, highlighting the model’s use for personalized biomarker prediction and sensitive-region identification.

\subsection{Individual-level simulations and predictions}
\begin{figure*}[!t]
    \centering
    \includegraphics[width=0.85\textwidth]{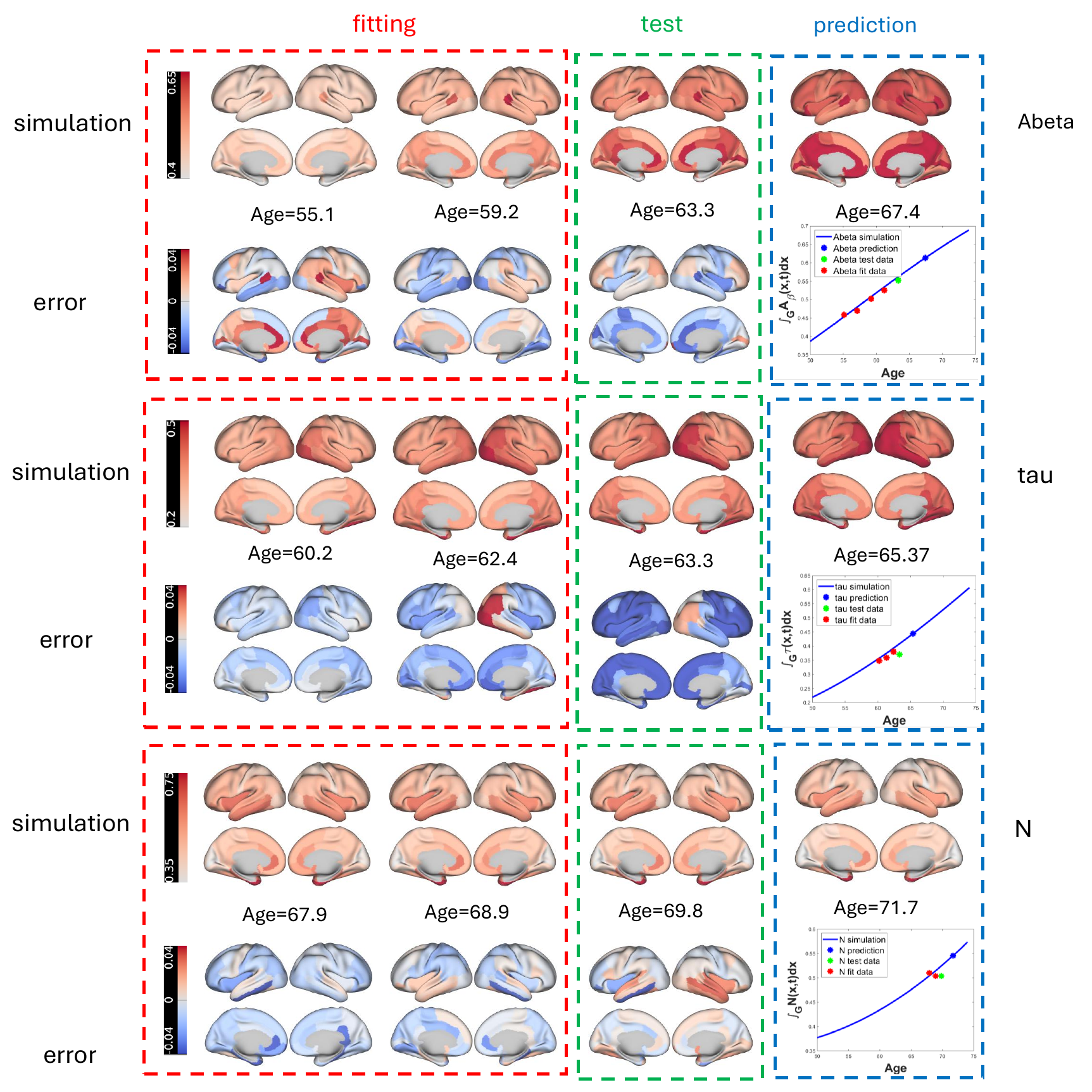}
    \caption{Model simulations of three spatially dependent biomarkers (A$\beta$, tau, N) for three individuals. Training fits, test evaluations, and future predictions are color-coded as red, green, and blue, respectively. Regional curves show accumulated biomarker levels across 68 brain regions.}
    \label{fig:Abeta_tau_N}
\end{figure*}

Fig.~\ref{fig:Abeta_tau_N} demonstrates the model’s performance for three representative individuals. The model achieves prediction accuracies of $96.15\%$, $93.97\%$, and $93.35\%$ for A$\beta$, tau, and N, respectively. Training fits (red rectangles), test evaluations (green), and future predictions (blue) are shown, with curves representing accumulated biomarker levels across 68 brain regions. By visualizing only two representative fitting points per biomarker, we simplify interpretation while preserving the spatial distribution of pathology. These results illustrate the model’s capability to forecast individualized biomarker trajectories, providing actionable information for clinicians to anticipate disease progression and personalize interventions.

\subsection{Population-level evaluation \rev{for one time point prediction }}

To assess generalizability, we evaluated performance across all subjects (Table~\ref{tab:accuracy_summary}). The proposed nonhomogeneous (NH) PDE model, which allows region-specific parameters to capture spatial heterogeneity, consistently outperforms the homogeneous (H) benchmark in both training and testing. On the test set, the NH model achieves higher mean accuracy for every outcome, with average gains of 1.87\% (A$\beta$), 1.44\% (tau), 4.40\% (N), and 4.88\% (C). In addition, the NH model shows smaller standard deviations, indicating more stable and robust predictions across subjects.

\begin{figure}[!t]
    \centering
    \includegraphics[width=0.8\textwidth]{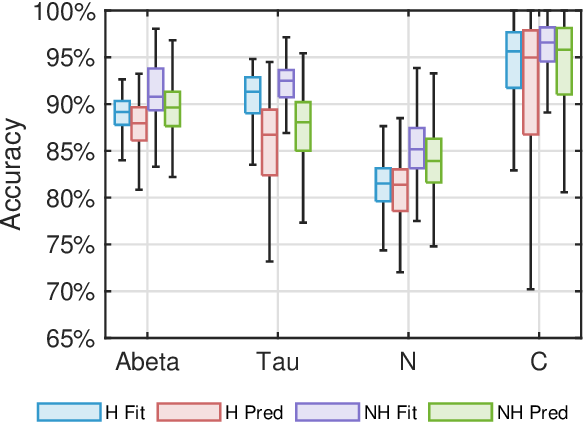}
 \caption{Boxplots of model accuracy across all subjects for H and NH models (interquartile range and median shown).}
    \label{fig:model-popu}
\end{figure}

\begin{figure}[!t]
    \centering
    \includegraphics[width=0.9\columnwidth]{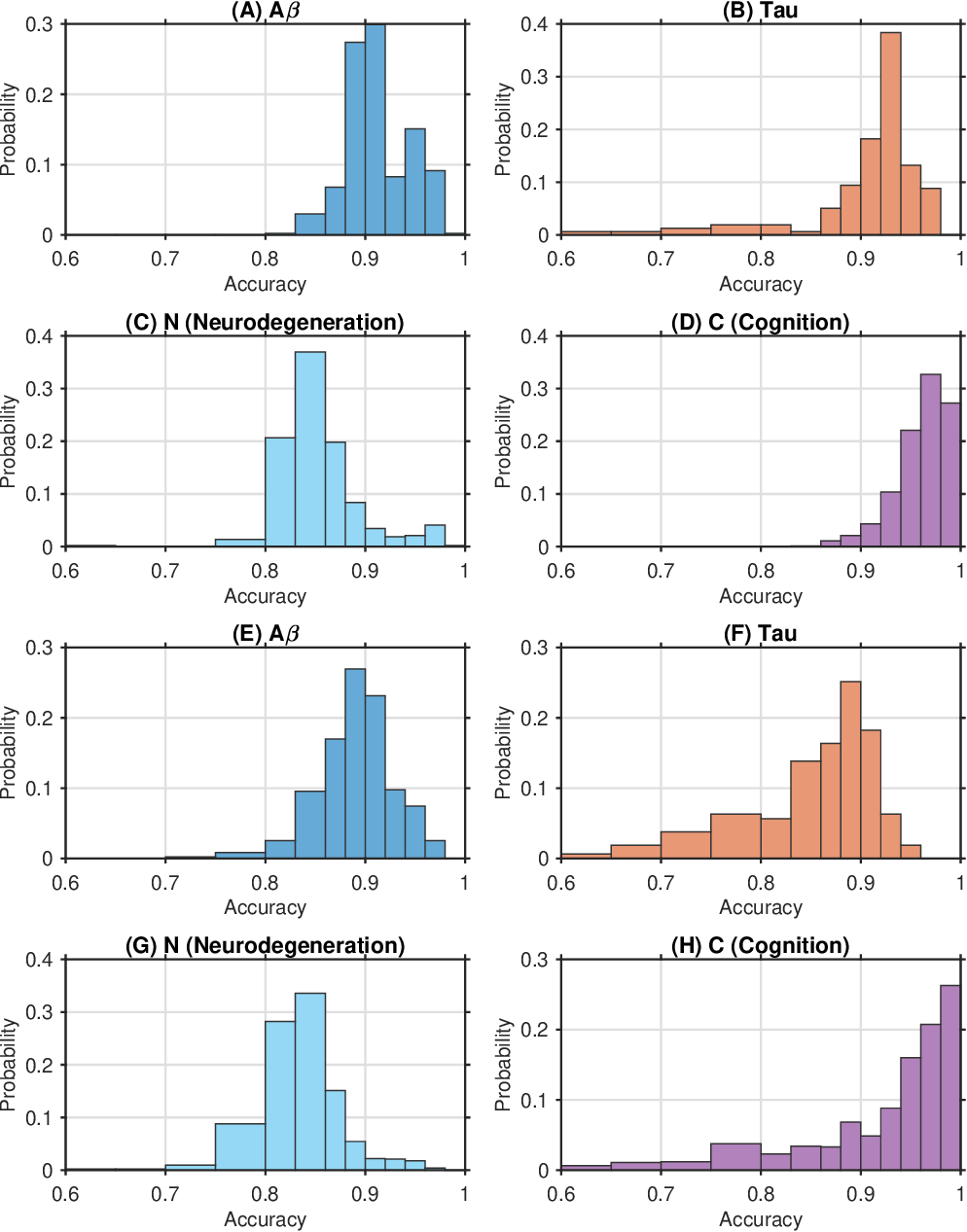} 
    \caption{\rev{Distributions of fitting accuracies (top) and prediction accuracies (bottom) of the nonhomogeneous (NH) model across all subjects for the four biomarkers: A$\beta$, tau, neurodegeneration (N), and cognition (C). Each histogram shows the empirical distribution of subject-specific accuracies for the corresponding biomarker. The y-axis represents the probability (normalized frequency) of observations within each accuracy bin.} }
    \label{fig:model-popu1}
\end{figure}

Boxplots (Fig.~\ref{fig:model-popu}) and histograms (Fig.~\ref{fig:model-popu1}) further confirm that most subjects achieve over 80\% accuracy across biomarkers under the NH model. Together, these results support the utility of incorporating regional heterogeneity for reliable, personalized predictions at the population level.
\begin{table}[!t]
\centering
\caption{Summary reported as (median, mean, std) for each biomarker for homogeneous (H) model and nonhomogeneous (NH) model \rev{for one-time-point prediction}.}
\label{tab:accuracy_summary}
\setlength{\tabcolsep}{6pt}    
\renewcommand{\arraystretch}{1.12}

\begin{tabular}{lcc}
\toprule
\multicolumn{3}{c}{\textbf{Homogeneous (H) model}} \\
\midrule
\multicolumn{1}{c}{\textbf{Biomarker}} & \textbf{Fit (H) (\%)} & \textbf{Test (H) (\%)} \\
\midrule
Abeta & (89.15, 88.87, 1.98) & (87.95, 87.61, 2.87) \\
tau   & (91.32, 89.48, 5.57) & (86.71, 85.07, 6.16) \\
N     & (81.51, 80.03, 8.18) & (81.38, 79.30, 9.09) \\
C     & (95.63, 91.33, 12.31) & (94.98, 88.26, 16.87) \\
\midrule
\multicolumn{3}{c}{\textbf{Nonhomogeneous (NH) model}} \\
\midrule
\textbf{Biomarker} & \textbf{Fit (NH) (\%)} & \textbf{Test (NH) (\%)} \\
\midrule
Abeta & (90.78, 91.33, 3.05) & (89.63, 89.48, 3.44) \\
tau   & (92.49, 91.36, 5.01) & (88.06, 86.51, 5.62) \\
N     & (85.16, 85.49, 5.80) & (83.92, 83.70, 5.82) \\
C     & (96.58, 95.80, 3.52) & (95.81, 93.14, 7.59) \\
\bottomrule
\end{tabular}
\end{table}

\rev{\subsection{Population-level evaluation for long term prediction}
Based on the two-step prediction results summarized in Table~\ref{tab:accuracy_summary2}, we further evaluated the model's multi-step forecasting ability by training on the first $N-2$ time points and predicting the last two visits. This two-step-ahead validation provides a more stringent test of long-term predictive performance compared to the one-step-ahead setting. Consistent with the one-step results, the nonhomogeneous (NH) model consistently outperforms the homogeneous (H) benchmark across all biomarkers in both fitting and testing. On the test set, the NH model achieves higher mean accuracy for every biomarker, with average gains of 1.63\% for A$\beta$, 15.46\% for tau, 1.43\% for N, and 2.11\% for C. Notably, the improvement for tau is substantial, increasing from 72.20\% (H) to 87.66\% (NH), indicating that region-specific parameters are critical for capturing the heterogeneous progression of tau pathology. For N and C, the NH model also demonstrates more stable predictions, as reflected by smaller standard deviations (2.30\% vs. 2.63\% for N; 5.99\% vs. 8.87\% for C on the test set). The two-step prediction results are consistent with the one-step findings, confirming that the NH model's superiority is not limited to short-term forecasting but generalizes to longer-term trajectory prediction. This multi-step validation directly addresses the concern regarding overestimation of predictive performance, demonstrating that the proposed model maintains robust accuracy even when tasked with predicting multiple future time points. The consistency across both evaluation settings (one-step and two-step) provides strong evidence for the model's ability to capture underlying disease progression dynamics and its potential utility for clinical trajectory forecasting.}

\begin{table}[!t]
\centering
\begin{revtableblock}
\caption{\rev{Summary reported as (median, mean, std) for each biomarker for homogeneous (H) model and nonhomogeneous (NH) model for two-time-point prediction.} }
\label{tab:accuracy_summary2}
\setlength{\tabcolsep}{6pt}
\renewcommand{\arraystretch}{1.12}
\begin{tabular}{lcc}
\toprule
\multicolumn{3}{c}{\textbf{Homogeneous (H) model}} \\
\midrule
\multicolumn{1}{c}{\textbf{Biomarker}} & \textbf{Fit (H) (\%)} & \textbf{Test (H) (\%)} \\
\midrule
Abeta & (89.25, 89.25, 1.98) & (88.37, 88.01, 2.05) \\
tau & (70.15, 71.44, 19.19) & (71.91, 72.20, 16.84) \\
N & (78.20, 77.84, 3.97) & (80.25, 79.94, 2.63) \\
C & (96.37, 96.74, 1.25) & (96.51, 93.01, 8.87) \\
\midrule
\multicolumn{3}{c}{\textbf{Nonhomogeneous (NH) model}} \\
\midrule
\textbf{Biomarker} & \textbf{Fit (NH) (\%)} & \textbf{Test (NH) (\%)} \\
\midrule
Abeta & (91.19, 91.30, 2.63) & (89.31, 89.64, 1.99) \\
tau & (93.53, 93.41, 1.82) & (87.57, 87.66, 4.34) \\
N & (80.45, 80.51, 1.42) & (81.31, 81.37, 2.30) \\
C & (96.37, 96.74, 1.26) & (97.38, 95.12, 5.99) \\
\bottomrule
\end{tabular}
\end{revtableblock}
\end{table}

\rev{\subsection{Stability analysis of thresholds for creating functional \revtwo{connectivity} network}
We further assessed the sensitivity of the model to the choice of FC threshold used for network construction. As summarized in Table~\ref{tab:threshold-fc}, the prediction accuracy remained consistently high across all thresholds, with median test accuracies ranging from 88.76\% to 90.15\%. Notably, the standard deviations of test accuracies decreased substantially at thresholds of 0.75 and 0.80 (from 7.51\% to approximately 2\%), indicating stable and reliable predictions. These findings demonstrate that the proposed nonhomogeneous model is highly robust to the specific FC threshold selection, maintaining strong predictive performance for A$\beta$ deposition across a reasonable range of network configurations.}

\begin{table}[h]
\centering 
\begin{revtableblock}
\caption{\rev{Summary statistics of fit and prediction accuracy of nonhomogeneous model for predicting A$\beta$ with different thresholds.}}
\label{tab:threshold-fc}
\begin{tabular}{l|cccccc}
\toprule
\textbf{Threshold} & \multicolumn{3}{c|}{\textbf{Fit (\%)}} & \multicolumn{3}{c}{\textbf{Test (\%)}} \\
 & Median & Mean & Std & Median & Mean & Std \\
\midrule
0.70 & 89.25 & 87.73 & 5.65 & 88.76 & 86.37 & 7.51 \\
0.75 & 91.19 & 91.30 & 2.63 & 89.31 & 89.64 & 1.99 \\
0.80 & 91.25 & 91.41 & 2.16 & 90.15 & 89.86 & 2.70 \\
\bottomrule
\end{tabular}
\end{revtableblock}
\end{table}

\subsection{Impact of training strategies}
\begin{figure}[!t]
    \centering
    \includegraphics[width=0.9\columnwidth]{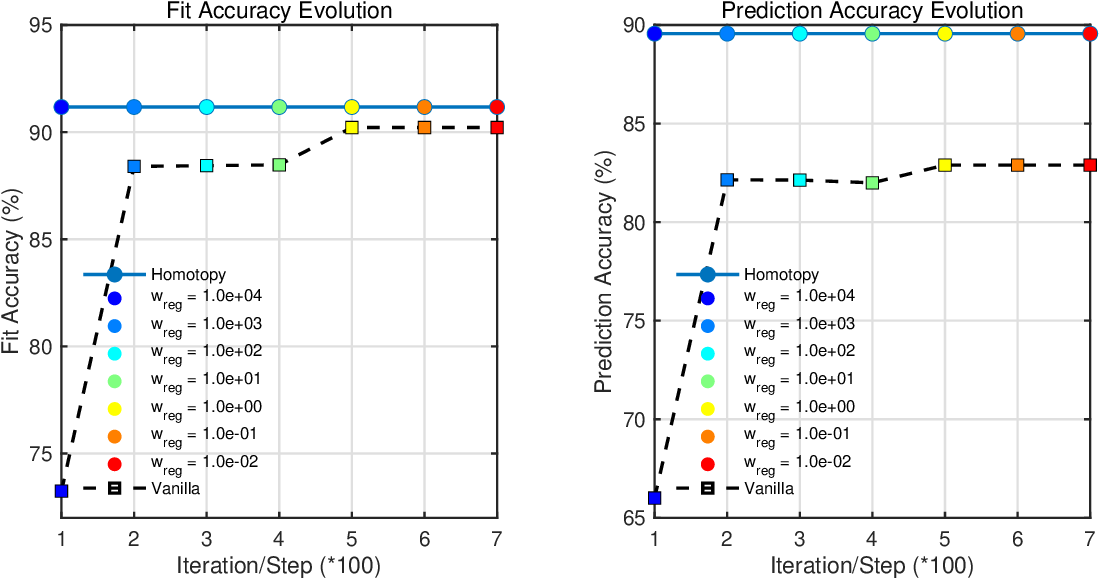}
    \caption{\rev{Comparison of homotopy regularization (solid lines with circles) and vanilla training (dashed lines with squares) across optimizer iterations. Homotopy training maintains consistently high fitting and prediction accuracy as the regularization weight is gradually reduced, whereas vanilla optimization exhibits substantially lower performance.}}
    \label{fig:homo-reg-svd}
\end{figure}

\rev{We next evaluated the impact of the proposed hierarchical structured training strategy with homotopy regularization technique on parameter estimation and predictive performance. As shown in Fig.~\ref{fig:homo-reg-svd}, homotopy regularization consistently outperforms conventional training across a range of regularization weights. Compared with vanilla optimization, the proposed strategy improves both fitting and prediction accuracy, particularly in the intermediate regularization regime, while maintaining stable performance throughout the continuation process. These results suggest that homotopy regularization effectively stabilizes optimization of the nonlinear PDE-constrained inverse problem and reduces susceptibility to poor local minima. Besides, as demonstrated in Table~\ref{tab:accuracy_summary} and Fig.~\ref{fig:model-popu}, the proposed hierarchical structured training strategy in Fig.~\ref{fig:HS} consistently yields equal or improved fitting and prediction accuracy compared with the preceding stage. This coarse-to-fine strategy enables efficient exploration of the high-dimensional parameter space while preserving information learned at earlier stages.}

\begin{figure}[!t]
    \centering
    \includegraphics[width=0.65\columnwidth]{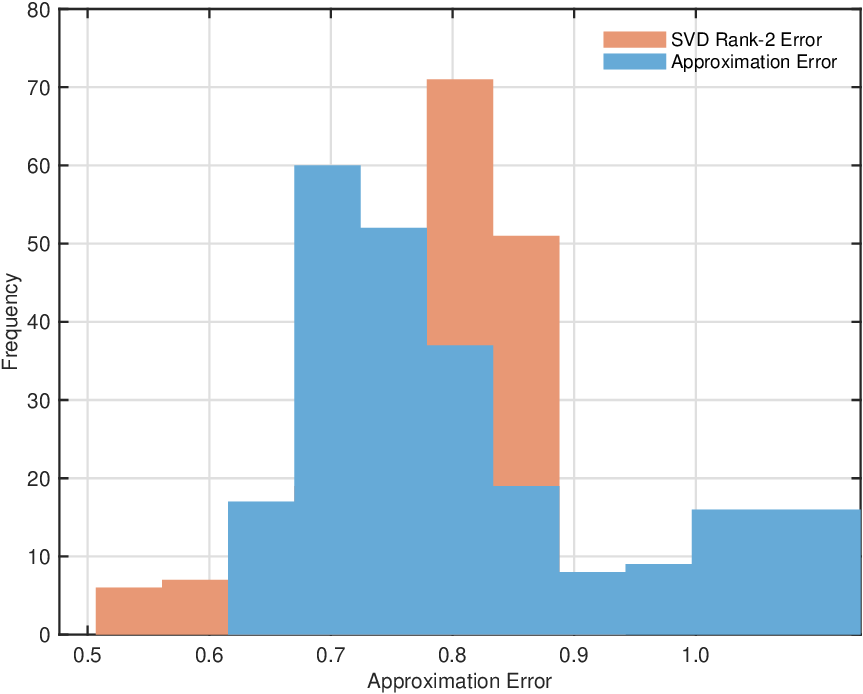}
    \caption{\rev{Distribution of relative approximation errors for patient-specific FC matrices. The proposed rank-2 latent representation achieves approximation accuracy comparable to the optimal rank-2 SVD reconstruction of the corresponding FC matrices, indicating that the dominant subject-specific connectivity structure can be captured using only a small number of latent parameters.}}
    \label{fig:homo-reg-svd1}
\end{figure}

\rev{\subsection{Effectiveness of low-rank approximation \revtwo{for patient-specific functional connectivity network}}
We further evaluated the effectiveness of the proposed low-rank representation for personalized functional connectivity estimation. Figure~\ref{fig:homo-reg-svd1} compares the approximation error of the learned patient-specific FC matrices with that of the optimal rank-2 SVD approximation. Despite being inferred indirectly from biomarker observations rather than FC measurements, the proposed representation achieves an error distribution comparable to the corresponding rank-2 SVD reconstruction. Since rank-2 SVD provides the optimal low-rank approximation in the least-squares sense, these results suggest that the learned representation captures the dominant subject-specific connectivity patterns relevant to disease progression.}

\revtwo{
To rigorously isolate and quantify the contribution of patient-specific functional connectivity (FC) to predictive performance, we conducted a controlled ablation study. Specifically, we held all learned nonhomogeneous model parameters and initial conditions fixed and compared two FC configurations: (i) population-average FC, implemented by setting $\bu=\bv=0$, and (ii) patient-specific FC, in which $\bu$ and $\bv$ were optimized from individual neuroimaging data. By keeping all other model components identical, this design ensures that any difference in predictive performance can be attributed specifically to the incorporation of personalized connectivity information.

Using the biomarker A$\beta$ as a representative case, Table~\ref{tab:Lp_Lp_uv} reports the mean$\pm$standard deviation of test accuracy and the maximum individual-level improvement achieved by incorporating patient-specific FC.
At the population level, incorporating patient-specific FC results in only a modest improvement in predictive accuracy, with the mean test accuracy increasing from $89.25\pm3.51\%$ to $89.48\pm3.44\%$. This relatively small change indicates that population-average FC provides a strong baseline for A$\beta$ prediction at the cohort level. Nevertheless, the individual-level analysis reveals substantially larger benefits for specific patients, with the maximum improvement in test accuracy reaching 9.73 percentage points. These results suggest that the predictive value of patient-specific FC is heterogeneous across individuals: while population-average connectivity may adequately represent patients whose functional connectivity patterns are close to the population norm, personalized FC can provide substantial additional predictive information for patients whose connectivity patterns deviate more substantially from the population average. Thus, although patient-specific FC does not substantially improve aggregate predictive performance, it can yield meaningful individualized gains for a subset of patients.
}

\begin{table}[!t]
\centering
\begin{revtableblocktwo}
\caption{\revtwo{Comparison of prediction accuracy for A$\beta$ with and without patient-specific $\bu,\bv$ in the non-homogeneous model. Accuracy is reported as mean$\pm$std, together with the maximum individual-level improvement in accuracy}}
\label{tab:Lp_Lp_uv}
\setlength{\tabcolsep}{6pt}
\renewcommand{\arraystretch}{1.12}
\begin{tabular}{lccc}
\toprule
\textbf{Metric} & \textbf{Fit (\%)} & \textbf{Test (\%)} \\
\midrule
without $\bu,\bv$  & 91.08$\pm$ 3.06, baseline & 89.25$\pm$3.51, baseline \\
\midrule
with $\bu,\bv$ & 91.33$\pm$3.05, 8.36 & 89.48$\pm$ 3.44, 9.73 \\
\bottomrule
\end{tabular}
\end{revtableblocktwo}
\end{table}

%

\subsection{Two-level sensitivity analysis}
\begin{figure*}[!t]
    \centering
    \includegraphics[width=0.9\linewidth]{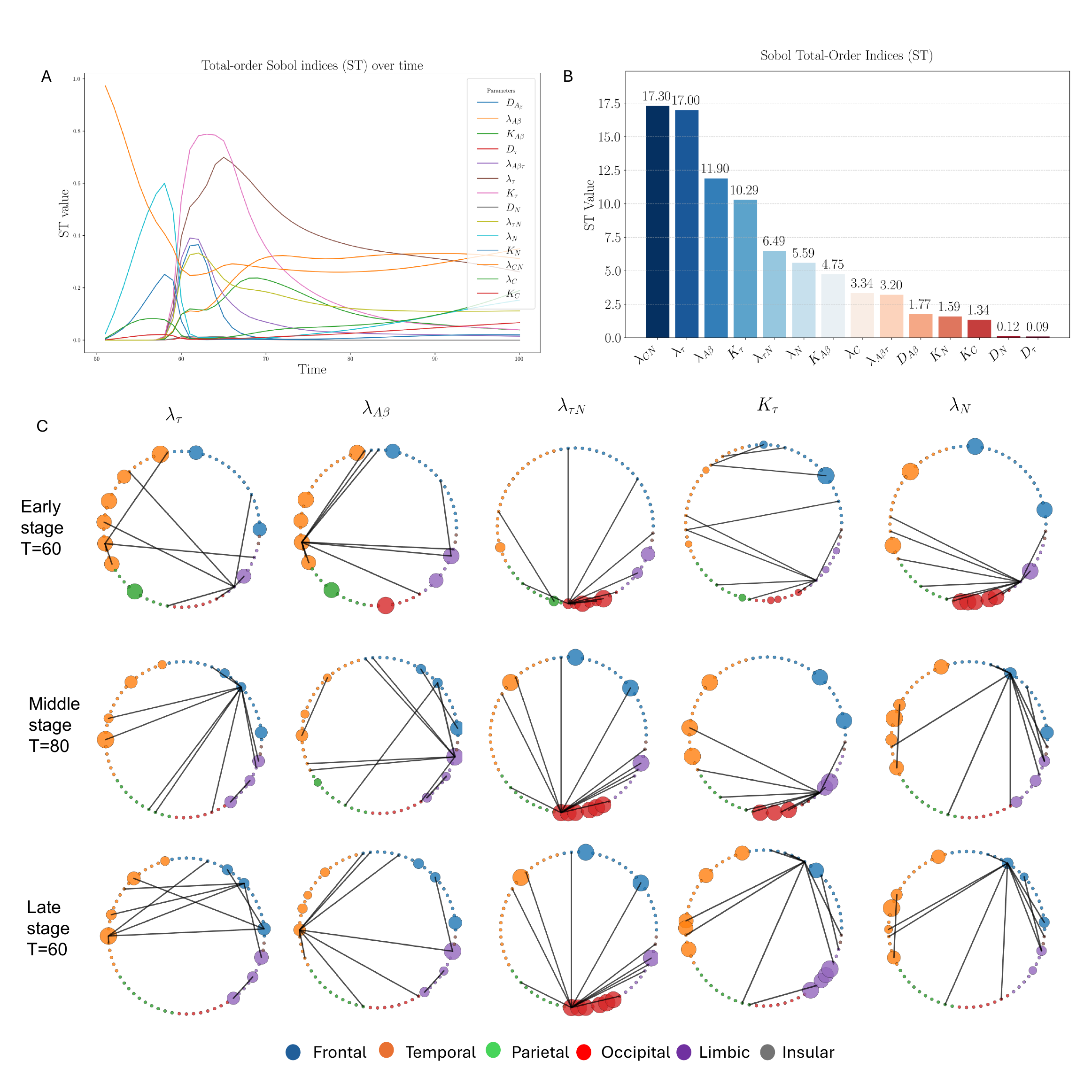}
    \caption{A: Dynamics of total sensitivity indices for 14 model parameters over age. B: Importance bar plot of sorted model parameters based on total sensitivity indices over age. C: The 68 Desikan–Killiany (DK) regions are grouped into six anatomical lobes. The second-level sensitivity analysis of key region-specific parameters is shown across three time stages ($t = 60$, $80$, and $100$). The first-order Sobol index ($S_1$), represented by the circle radius, quantifies each lobe’s direct contribution to the model variance, while the second-order index ($S_2$), encoded by edge width, captures the strength of pairwise interactions between lobes. During computations, the model’s initial condition is set to the population mean at $t = 50$, and the parameter ranges are defined by the minimum and maximum values of each parameter across the 68 regions. Each second-level sensitivity analysis is performed for a single region-specific parameter, with all other parameters fixed at the mean of the optimized values across all subjects.}
    \label{fig:sen}
\end{figure*}

We performed a two-level sensitivity analysis to identify critical model parameters and brain regions. First, assuming homogeneous parameters across all 68 DK regions, total sensitivity indices were calculated over age (Fig.~\ref{fig:sen}A), identifying six highly influential parameters: $\lambda_{CN}$, $\lambda_{\tau}$, $\lambda_{A_\beta}$, $K_{\tau}$, $\lambda_{\tau N}$, and $\lambda_{N}$ (Fig.~\ref{fig:sen}B). Second, we analyzed the top five region-specific parameters to pinpoint sensitive brain regions (Fig.~\ref{fig:sen}C), providing a quantitative basis for targeted monitoring and intervention.

Stage-wise sensitivity analysis reveals a dynamic network propagation mechanism: at early stages ($t=60$), first-order sensitivity indices ($S_1$) are relatively uniform, with modest cross-lobe interactions ($S_2$). By mid-stage ($t=80$), temporal–frontal, temporal–parietal, and temporal–limbic interactions dominate, while at late stage ($t=100$), inter-lobar interactions drive model sensitivity, with frontal–temporal and frontal–limbic couplings particularly prominent. These findings quantitatively support a network-based mechanism of disease spread, consistent with connectome-driven hypotheses.

Table~\ref{tab:sen-S1} summarizes top DK regions for each biomarker and disease stage. The temporal and frontal lobes are most affected, with involvement intensifying over disease progression, whereas the insular lobe remains minimally affected. Aggregated across biomarkers, frontal and temporal lobes show increasing involvement from early to late stages, confirming that network hubs play a central role in AD progression~\cite{murray2015clinicopathologic}.
 
\begin{table*}[!t]
\scriptsize
\raggedright
\caption{Top DK regions by biomarkers and three disease stages with lobe distribution. \rev{F: Frontal, T: Temporal, P: Parietal, O: Occipital, L: Limbic, I: Insular}}
\label{tab:sen-S1}
\begin{tabular}{llll}
\toprule
 & \textbf{Early Stage (t=60)} & \textbf{Middle Stage (t=80)} & \textbf{Late Stage (t=100)} \\
\midrule

\multirow{10}{*}{\textbf{$A_\beta$}}
& LH Posteriorcingulate (L) & LH Medialorbitofrontal (P) & LH Temporalpole (T) \\
& RH Temporalpole (T) & LH Parahippocampal (T) & RH Bankssts (L) \\
& RH Postcentral (T) & RH Cuneus (T) & LH Parahippocampal (T) \\
& RH Insula (T) & RH Bankssts (L) & RH Cuneus (T) \\
& RH Parsorbitalis (F) & LH Caudalanteriorcingulate (L) & LH Caudalanteriorcingulate (L) \\
& LH Transversetemporal (P) & RH Precuneus (L) & RH Precuneus (L) \\
& RH Posteriorcingulate (L) & LH Frontalpole (F) & LH Frontalpole (F) \\
& LH Unknown (T) & RH Paracentral (T) & RH Paracentral (T) \\
& LH Postcentral (O) & RH Entorhinal (F) & RH Entorhinal (F) \\
& LH Bankssts (T) & LH Rostralanteriorcingulate (F) & LH Rostralanteriorcingulate (F) \\
\hline
\textbf{Lobe}
& \makecell{F:1, T:5, P:1, O:1, L:2}
& \makecell{F:3, T:3, P:1, O:0, L:3}
& \makecell{F:3, T:4, P:0, O:0, L:3} \\
\hline
\multirow{10}{*}{\textbf{$\tau$}}
& LH Caudalmiddlefrontal (F) & LH Unknown (T) & RH Cuneus (T) \\
& RH Superiorfrontal (T) & RH Entorhinal (F) & LH Bankssts (T) \\
& LH Pericalcarine (T) & LH Bankssts (T) & RH Paracentral (T) \\
& RH Insula (T) & RH Paracentral (T) & LH Caudalanteriorcingulate (L) \\
& RH Posteriorcingulate (L) & LH Caudalanteriorcingulate (L) & RH Bankssts (L) \\
& LH Precuneus (L) & LH Rostralanteriorcingulate (F) & LH Rostralanteriorcingulate (F) \\
& LH Postcentral (O) & LH Parahippocampal (T) & LH Parahippocampal (T) \\
& LH Precentral (F) & RH Bankssts (L) & RH Entorhinal (F) \\
& RH Middletemporal (P) & LH Frontalpole (F) & LH Frontalpole (F) \\
& RH Precentral (F) & RH Precuneus (L) & RH Precuneus (L) \\
\hline
\textbf{Lobe}
& \makecell{F:3, T:3, P:1, O:1, L:2}
& \makecell{F:3, T:4, P:0, O:0, L:3}
& \makecell{F:3, T:4, P:0, O:0, L:3} \\
\hline
\multirow{10}{*}{\textbf{$N$}}
& RH Middletemporal (P) & RH Lateralorbitofrontal (O) & RH Superiorfrontal (T) \\
& RH Superiorfrontal (T) & RH Middletemporal (P) & LH Cuneus (O) \\
& RH Caudalmiddlefrontal (O) & LH Precuneus (L) & LH Pericalcarine (T) \\
& LH Pericalcarine (T) & LH Postcentral (O) & RH Caudalmiddlefrontal (O) \\
& LH Postcentral (O) & LH Lingual (O) & LH Frontalpole (F) \\
& LH Lingual (O) & RH Superiorfrontal (T) & LH Entorhinal (T) \\
& LH Cuneus (O) & LH Cuneus (O) & RH Entorhinal (F) \\
& RH Lateralorbitofrontal (O) & LH Paracentral (P) & RH Cuneus (T) \\
& LH Precuneus (L) & LH Pericalcarine (T) & RH Supramarginal (T) \\
& RH Posteriorcingulate (L) & RH Caudalmiddlefrontal (O) & LH Temporalpole (T) \\
\hline
\textbf{Lobe}
& \makecell{F:0, T:2, P:1, O:5, L:2}
& \makecell{F:0, T:2, P:2, O:5, L:1}
& \makecell{F:2, T:6, P:0, O:2, L:0} \\
\hline
\textbf{Total}
& \makecell{F:4, T:10, P:3, O:7, L:6}
& \makecell{F:6, T:9, P:3, O:5, L:7}
& \makecell{F:8, T:14, P:0, O:2, L:6} \\
\bottomrule
\end{tabular}
\end{table*}

\section{Discussion}
\label{sec:discussion}
We present a region-specific, spatiotemporal model of AD progression that integrates biomarker dynamics within the ATN framework. By coupling PDEs on the brain functional-connectivity network with a two-level sensitivity analysis, we identified both key parameters and spatiotemporal patterns that govern the cascade from amyloid-$\beta$ (A$\beta$) deposition to tau ($\tau$) aggregation and subsequent neurodegeneration (N). This framework provides a mechanistic, patient-specific approach to capture the dynamics of AD biomarkers and their propagation across the brain.

\subsection{Model performance and predictive reliability}

Leveraging multimodal neuroimaging data from the ADNI cohort, we parametrized the PDE model to construct individualized, region-specific biomarker trajectories. Unlike conventional AD criteria, which often overlook interindividual heterogeneity and focus on convergent phenotypes \cite{korczyn2024alzheimer}, our approach emphasizes personalized modeling as a pathway toward precision medicine.

The model demonstrated high accuracy in both fitting and prediction for A$\beta$, $\tau$, N, and cognitive decline (Fig.~\ref{fig:model-popu}, Table~\ref{tab:accuracy_summary}), with consistently low variance between training and testing errors. Homotopy-regularized training improved convergence and optimization stability compared to standard training (Fig.~\ref{fig:homo-reg-svd}), mitigating overfitting in this high-dimensional parameter space. Furthermore, low-rank approximations of subject-specific FC matrices achieved comparable accuracy to SVD rank-2 approximations of true FC matrices, demonstrating that computational efficiency can be achieved without compromising predictive performance. {\rev{These results also support the use of the rank-two parameterization in \eqref{eq:adjacency-matrix}. The low-rank representation reflects a deliberate trade-off between model flexibility and parameter identifiability. While higher-rank parameterizations could potentially capture more complex subject-specific connectivity patterns, they would also introduce substantially more parameters relative to the limited amount of longitudinal rs-fMRI data available for each subject. The observed agreement between the learned FC matrices and the corresponding rank-two SVD approximations suggests that the proposed representation captures the dominant patient-specific connectivity variations while maintaining computational efficiency and robustness.} \revtwo{The ablation study of $\bu,\bv$ further demonstrates that patient-specific FC provides a tangible, albeit modest, improvement over population-average FC. While the cohort-level gain is limited, the substantial individual-level improvements indicate that patient-specific $\bu$ and $\bv$ can provide meaningful additional predictive information for certain patients, particularly those whose connectivity patterns deviate from the population average. The modest aggregate improvement may partly reflect limitations in both the available neuroimaging data and the current low-rank parameterization, where the rank-two update may be too restrictive to fully capture the complexity of individual-specific FC patterns. With larger, higher-quality, and more longitudinally sampled datasets, together with more expressive personalized FC parameterizations, the benefits of patient-specific connectivity may become more pronounced.}} Collectively, these results validate both the robustness and the predictive reliability of the model, supporting its use for patient-specific forecasting.

\subsection{Regional vulnerabilities and biological interpretation}

The sensitivity analysis elucidated key parameters ($\lambda_{CN}$, $\lambda_{\tau}$, $\lambda_{A_\beta}$, $K_{\tau}$, $\lambda_{\tau N}$, $\lambda_{N}$) that drive biomarker dynamics and highlighted spatiotemporal patterns of vulnerability \cite{vogel2020characterizing}. Across disease stages, the temporal lobe consistently emerged as the earliest and most affected region, with frontal lobe involvement increasing in mid-to-late stages, whereas the insular lobe remained minimally affected and parietal/occipital lobes were relatively spared (Table~\ref{tab:sen-S1}). These patterns align with established neuropathological observations \cite{braak1991neuropathological, thal2002phases, planche2022mri}, wherein tau pathology initiates in the entorhinal and hippocampal cortices before propagating to temporal and frontal association areas. Imaging studies \cite{singh2006spatial, du2007different} similarly report cortical thinning beginning in medial temporal regions and progressing anteriorly with disease severity. Our findings reinforce the temporal lobe as a neurodegenerative epicenter and the frontal lobe as a late-stage amplifier.

\subsection{Amyloid, tau, and neurodegeneration trajectories}

The model recapitulates empirical biomarker propagation patterns observed in PET and MRI studies. Early A$\beta$ accumulation was predicted in temporobasal and frontomedial cortices, followed by expansion to frontal and limbic regions \cite{johnson2016tau, collij2022spatial, nestor2006declarative}. Tau propagation persisted in temporal and frontal lobes across stages, consistent with functional-pathway-based stepwise spread \cite{ossenkoppele2016tau, berron2021early}. Neurodegeneration, reflected in cortical thinning, intensified primarily in frontal and temporal lobes during late stages, in agreement with longitudinal MRI studies \cite{scahill2002mapping, desikan2008mri}. These trajectories capture the hallmark spatial hierarchy of AD progression: limbic–temporal initiation, frontal expansion, and relative parietal/occipital sparing.

\subsection{Network interactions and clinical implications}

Second-order sensitivity analysis revealed that inter-lobe interactions dominate over intra-lobe effects, suggesting that AD spreads as a coordinated network disruption rather than isolated regional atrophy. This observation supports connectome-driven propagation frameworks \cite{murray2015clinicopathologic, torok2025directionality, bougacha2025contributions, torok2023connectome, abdelnour2022advances}, highlighting the role of functional connectivity in mediating cross-lobar pathology. By leveraging functional rather than purely structural connectivity, our model captures dynamic interactions that may underlie symptom evolution: frontal–temporal interplay could correspond to transitions from memory to executive dysfunction, while late insular involvement may relate to emotional and interoceptive deficits. These insights emphasize the potential for network-targeted therapeutic strategies that aim to preserve functional resilience rather than focusing solely on single regions.

\subsection{Limitations and future directions}

\rev{Several limitations of the current framework should be acknowledged.

First, the proposed biomarker cascade model represents a simplified description of AD progression. The current formulation assumes a predominantly sequential pathway from amyloid-beta accumulation to tau pathology, neurodegeneration, and cognitive decline, consistent with the amyloid cascade hypothesis and ATN framework which has been adopted in numerous mathematical and computational models of disease progression \cite{hao2025optimal,petrella2024personalized,zheng2022data} . However, accumulating evidence suggests bidirectional interactions and parallel pathological processes involving tau, neuroinflammation, vascular dysfunction, and metabolic alterations\cite{zhou2025accelerating, shankar2008amyloid}. Direct amyloid-mediated neurotoxicity, biomarker feedback loops, and neuroinflammatory mechanisms are not explicitly represented in the current model. Future extensions will incorporate additional state variables and interaction pathways as richer longitudinal multimodal datasets become available.

Second, the model assumes a fixed patient-specific graph Laplacian throughout the observation period. Although functional connectivity evolves during aging and disease progression, reliable estimation of dynamic subject-specific connectivity requires denser longitudinal rs-fMRI data than are currently available. Consequently, the learned graph Laplacian should be interpreted as an individualized network substrate capturing dominant disease-relevant coupling patterns. Future work will investigate time-dependent graph Laplacians and dynamic network reorganization.

\revtwo{Third, another limitation concerns our incorporation of FC into the diffusion term. Our formulation is motivated by the graph Laplacian reaction--diffusion framework, in which FC naturally represents inter-regional coupling and governs the propagation of pathology through the diffusion process across brain regions. Nevertheless, alternative formulations are also possible. For example, FC may additionally influence local biological processes and could be incorporated into the reaction term to modulate regional disease dynamics. Although the current formulation is consistent with a large body of mathematical modeling studies on graph-based reaction--diffusion systems and is further supported by our robustness analyses, we acknowledge that it is not the only biologically plausible modeling strategy. Future work should systematically compare diffusion-driven and reaction-driven FC formulations to better understand their biological implications and improve the mechanistic interpretability of the resulting model parameters.}

\revtwo{Fourth}, uncertainty in model predictions was not quantified. The current framework provides point estimates of subject-specific parameters and future biomarker trajectories without corresponding confidence intervals or prediction intervals, limiting our ability to assess the reliability of individualized forecasts. Future work will incorporate uncertainty quantification through Bayesian inference, probabilistic parameter estimation, and ensemble forecasting approaches, enabling more rigorous assessment of prediction confidence and improving the reliability of patient-specific forecasting.

Beyond these limitations, future work will integrate additional biological pathways, dynamic network evolution, uncertainty quantification, and richer multimodal datasets, including molecular, vascular, and neuroinflammatory biomarkers. Such extensions will further improve model personalization and provide a more comprehensive mechanistic framework for studying Alzheimer's disease progression and therapeutic interventions.
}

\section{Conclusion}
\label{sec:conclusion}
In this work, we developed a data-driven modeling framework to characterize the spatiotemporal progression of Alzheimer’s disease (AD) biomarkers and cognitive decline by integrating mathematical modeling, AD pathophysiology, neuroimaging data, graph theory, and sensitivity analysis. Specifically, we proposed a novel spatiotemporal model formulated as a system of partial differential equations (PDEs) defined on the brain’s functional connectivity network, capable of describing the regional propagation of the three key AD biomarkers and cognitive decline.

Using multimodal neuroimaging data from the ADNI cohort, we parametrized the PDE model to construct patient-specific and region-specific biomarker trajectories, enabling individualized characterization of disease dynamics. Recognizing that current AD criteria often overlook interindividual heterogeneity—thus limiting therapeutic efficacy as drug development targets convergent phenotypes rather than patient-specific mechanisms \cite{korczyn2024alzheimer}—our approach emphasizes personalized modeling as a pathway toward precision medicine in AD.

To address the high-dimensional, non-convex optimization problem arising in parameter estimation, we implemented a hierarchical training strategy combined with a homotopy regularization scheme, which ensures robust and efficient calibration of the biomarker cascade model.

Overall, the proposed PDE-based network model successfully reproduces the hallmark spatial-temporal features of AD, including early temporal–limbic vulnerability, progressive frontal involvement, and relative parietal–occipital sparing. By bridging mathematical modeling with neurobiological evidence, this framework establishes a quantitative foundation for understanding region-specific disease progression and offers a principled platform for designing personalized, network-aware therapeutic interventions in Alzheimer’s disease.

\section{Methods}
\label{sec:methods}
\subsection{Data description}
We accessed the multimodal data from the ADNI website following approval of our data use application (http://adni.loni.usc.edu/). The files titled “UC Berkeley - AV45 analysis [ADNI1,GO,2,3] (version:2020-05-12)” and  “UC Berkeley - AV1451 analysis [ADNI1,GO,2,3] (version:2022-04-26)” compiled by ADNI were utilized to obtain A$\beta$-PET and tau-PET regional standardized uptake value ratios (SUVRs), Regional SUVRs were determined by dividing the standardized uptake values (SUVs) of the target regions by the SUV of the whole cerebellum, which served as the reference region because of its low specific binding and consistent uptake across the study population. 

In this study, we quantify neurodegeneration using cortical thickness, a well-established MRI-based biomarker that reflects neuronal loss and structural atrophy in AD. This approach follows previous studies demonstrating that cortical thinning is strongly associated with AD pathology and predicts future cognitive decline\cite{dickerson2012mri,dickerson2011alzheimer,harrison2021distinct,mehta2024early, racine2018personalized, keuss2024rates}. 
Specifically, reduced cortical thickness in AD-signature regions such as the medial temporal, inferior parietal, and posterior cingulate cortices has been shown to correlate with amyloid and tau burden as well as with disease progression. Cortical thickness data was obtained from ADNI as part of the “UCSF-Cross-Sectional FreeSurfer (6.0) [ADNI3]” and “UCSF-Cross-Sectional FreeSurfer (5.1) [ADNI1, GO, 2]” datasets. 

Resting-state fMRI data were acquired on 3~Tesla MR scanners across multiple ADNI sites using a standardized ADNI imaging protocol (\url{https://adni.loni.usc.edu/data-samples/adni-data/neuroimaging/mri/mri-scanner-protocols/}). Each session included a high-resolution 3D T1-weighted MPRAGE scan for anatomical segmentation and spatial registration. Acquisition parameters for the T1-weighted scan were: field of view $= 208 \times 240 \times 256~\mathrm{mm}^3$, voxel size $= 1 \times 1 \times 1~\mathrm{mm}^3$, $\mathrm{TR} = 2300~\mathrm{ms}$, $\mathrm{TI} = 900~\mathrm{ms}$, TE = minimum full echo, and scan duration of approximately 6~min 20~s. Resting-state functional images were obtained using the ADNI-3 Basic gradient-echo EPI-BOLD sequence with the following parameters: field of view $= 220 \times 220 \times 163~\mathrm{mm}^3$, $\mathrm{TR} = 3000~\mathrm{ms}$, $\mathrm{TE} \approx 30~\mathrm{ms}$, and flip angle $= 90^\circ$. Each run lasted 10~min and produced approximately 200 volumes.       

Preprocessing was performed using a standard resting-state pipeline consistent with prior ADNI studies(han2021reduced, han2024global, mao2026global). The preprocessing steps included motion correction, skull stripping, spatial smoothing with a 4 mm FWHM Gaussian kernel, temporal filtering in the 0.01–0.1 Hz range, and alignment of functional images to the individual T1-weighted scan followed by transformation into MNI-152 standard space. To minimize transient magnetization effects and filtering boundary artifacts, the first five and last five volumes of each run were removed prior to analysis. Parcel-wise resting-state time series were then derived using the Desikan–Killiany–Tourville 68-region atlas\cite{DESIKAN2006968} by averaging the preprocessed BOLD signal within each cortical parcel. Functional connectivity (FC) data was derived by calculating Pearson's correlation coefficients between time series extracted from regions defined by the DKT 68 atlas\cite{DESIKAN2006968}, based on rsfMRI data. 

The Mini-Mental State Examination (MMSE) scores were obtained from “Mini-Mental State Examination (MMSE) [ADNI1,GO,2,3,4]”. In this study, we did not impose a strict requirement for each subject to have data available for all modalities mentioned. Instead, our inclusion criterion focused on ensuring that each subject had data from at least three separate visits for one of the key measurements: tau-PET, amyloid-beta (A$\beta$-PET), or cortical thickness.

The study included 1,891 subjects from ADNI who were classified as cognitively normal (CN), mild cognitive impairment (MCI), AD, or of unknown status. The details are shown in Table \ref{tab:data} and the demographic details are shown in Table \ref{tab:subjects_demog}.  Because the biomarker-cascade PDE system is sequentially driven (A$\beta \rightarrow \tau \rightarrow N \rightarrow C$), we estimate the model in a decoupled, stage-wise manner. Accordingly, the dataset is summarized in terms of \emph{task-specific cohorts}: for each sub-equation, we include all participants who have at least three longitudinal time points for the modality/modalities required to fit that sub-equation. Importantly, the resulting cohorts are \emph{not mutually exclusive}. For example, the ``A$\beta$ cohort'' is the union of all participants with longitudinal A$\beta$ PET data, regardless of whether $\tau$, $N$, and/or cognition ($C$, derived from MMSE) are also available. Table~\ref{tab:data} reports the effective sample sizes by diagnosis group for fitting each stage.

\begin{table*}[t]
\caption{Task-specific cohorts used in the sequential (decoupled) estimation of the biomarker-cascade PDE system.
For each sub-equation, we report the number of ADNI participants with at least three longitudinal time points
for the required modality/modalities. Rows correspond to \emph{effective sample sizes} for fitting each sub-equation
and are \emph{not mutually exclusive} (e.g., the A$\beta$ cohort is the union of all participants with A$\beta$ available,
regardless of whether $\tau$, $N$, and/or $C$ are also observed). CN: cognitively normal; MCI: mild cognitive impairment;
AD: Alzheimer’s disease.}
\label{tab:data}
\centering
\footnotesize
\setlength{\tabcolsep}{4pt}
\begin{tabular*}{\textwidth}{@{\extracolsep{\fill}} l p{5.1cm} r r r r r}
\toprule
Sub-equation & Required observed modalities & CN & MCI & AD & Unknown & Total \\
\midrule
A$\beta$ dynamics ($A_\beta$) 
& A$\beta$ 
& 142 & 140 & 3 & 186 & 471 \\
\midrule
$\tau$ dynamics ($\tau$)
& $\tau$ and A$\beta$ 
& 30 & 36 & 1 & 17 & 84 \\
& $\tau$ only (A$\beta$ not available) 
& 30 & 26 & 16 & 3 & 75 \\
\midrule
Neurodegeneration ($N$)
& A$\beta$ + $\tau$ + $N$ 
& 40 & 53 & 0 & 14 & 107 \\
& $\tau$ + $N$ only (A$\beta$ not available) 
& 19 & 9 & 4 & 0 & 32 \\
& A$\beta$ + $N$ only ($\tau$ not available) 
& 121 & 192 & 1 & 21 & 335 \\
& $N$ only (neither A$\beta$ nor $\tau$ available) 
& 87 & 264 & 109 & 20 & 480 \\
\midrule
Cognition ($C$)
& A$\beta$ + $\tau$ + $N$ + $C$ 
& 40 & 53 & 0 & 14 & 107 \\
& $\tau$ + $N$ + $C$ only (A$\beta$ not available) 
& 19 & 9 & 4 & 0 & 32 \\
& A$\beta$ + $N$ + $C$ only ($\tau$ not available) 
& 121 & 192 & 1 & 21 & 335 \\
& $N$ + $C$ only (neither A$\beta$ nor $\tau$ available) 
& 83 & 254 & 100 & 20 & 457 \\
\bottomrule
\end{tabular*}
\end{table*}

\begin{table*}[!t]
\centering
\caption{Subject characteristics after keeping participants with at least 3 longitudinal records for each modality. Values are counts unless stated otherwise.}
\label{tab:subjects_demog}
\footnotesize
\setlength{\tabcolsep}{4pt}
\renewcommand{\arraystretch}{1.12}
\begin{tabular*}{\textwidth}{@{\extracolsep{\fill}} l c c c c c}
\hline
 & \textbf{MMSE} & \textbf{A$\beta$} & \textbf{$\tau$} & $N$ \textbf{(12GO)} & $N$ \textbf{(ADNI3)} \\
\hline
Unique subjects ($N$) 
& 1889 & 471 & 159 & 806 & 148 \\
\hline
Gender (M/F/Missing) 
& 1027 / 862 / 0 & 236 / 235 / 0 & 73 / 86 / 0 & 436 / 370 / 0 & 69 / 79 / 0 \\
\hline
APOE genotype counts 
& \makecell[l]{2/2: 3\\2/3: 140\\2/4: 39\\3/3: 873\\3/4: 651\\4/4: 181\\Missing: 2}
& \makecell[l]{2/2: 1\\2/3: 47\\2/4: 8\\3/3: 260\\3/4: 129\\4/4: 26}
& \makecell[l]{2/3: 9\\2/4: 2\\3/3: 77\\3/4: 56\\4/4: 15}
& \makecell[l]{2/2: 1\\2/3: 66\\2/4: 14\\3/3: 369\\3/4: 279\\4/4: 77}
& \makecell[l]{2/3: 15\\2/4: 4\\3/3: 79\\3/4: 42\\4/4: 8} \\
\hline
Age, mean (SD) 
& 73.35 (7.50) & 72.76 (7.20) & 73.59 (7.60) & 73.42 (7.32) & 74.82 (7.03) \\
Age missing 
& 0 & 0 & 0 & 0 & 0 \\
\hline
\end{tabular*}
\end{table*}

\subsection{Data preparation}
We normalize data measurments across all subjects and brain subregions using:
\begin{equation}
    \by^i_{j} = \frac{\by^i_{j}}{\by^{ref}_{j}}
\end{equation}
where $i=1, 2, ..., N$ indexes the $i$-th subject, $j=1, 2, ..., 68$ denotes the $j$-th brain subregion. And $\by^{ref}_j$ is the reference value for $\by_j$ in the $j$-th region. $\by$ include amyloid-beta ($A_{\beta}$), tau ($\tau_\rho$),  neurodegeneration ($N$) and cognitive impairment $C$.  

To ensure uniform biomarker trajectories and cognitive decline trajectory aligned with disease progression (all increasing with severity), we define linear transformations on the normalized biomarkers and cognitive decline as follows: 
\begin{itemize}
    \item Amyloid $\beta$ and tau proteins: both $A_{\beta}$ and $\tau_\rho$ increase with disease progression, as measured in PET imaging. 
    \item Neurodegeneration: $N_{ct}$ quantified by cortical thickness which decreases as AD advances. We invert this trend by defining $N=1-N_{ct}$. 
    \item Cognitive decline: Measured via the Mini-Mental State Examination (MMSE) score $S$, which declines with worsening pathology. We reverse its directionality using $C=1-S$.
\end{itemize}
This normalization and transformation framework ensures consistent biomarker and cognitive decline behavior between subjects and subregions, simplifyingthe interpretation of the model. 

\subsection{Patient-specific spatiotemporal partial differential equations model}
\begin{figure}[!t]
 \centerline{\includegraphics[width=\columnwidth]{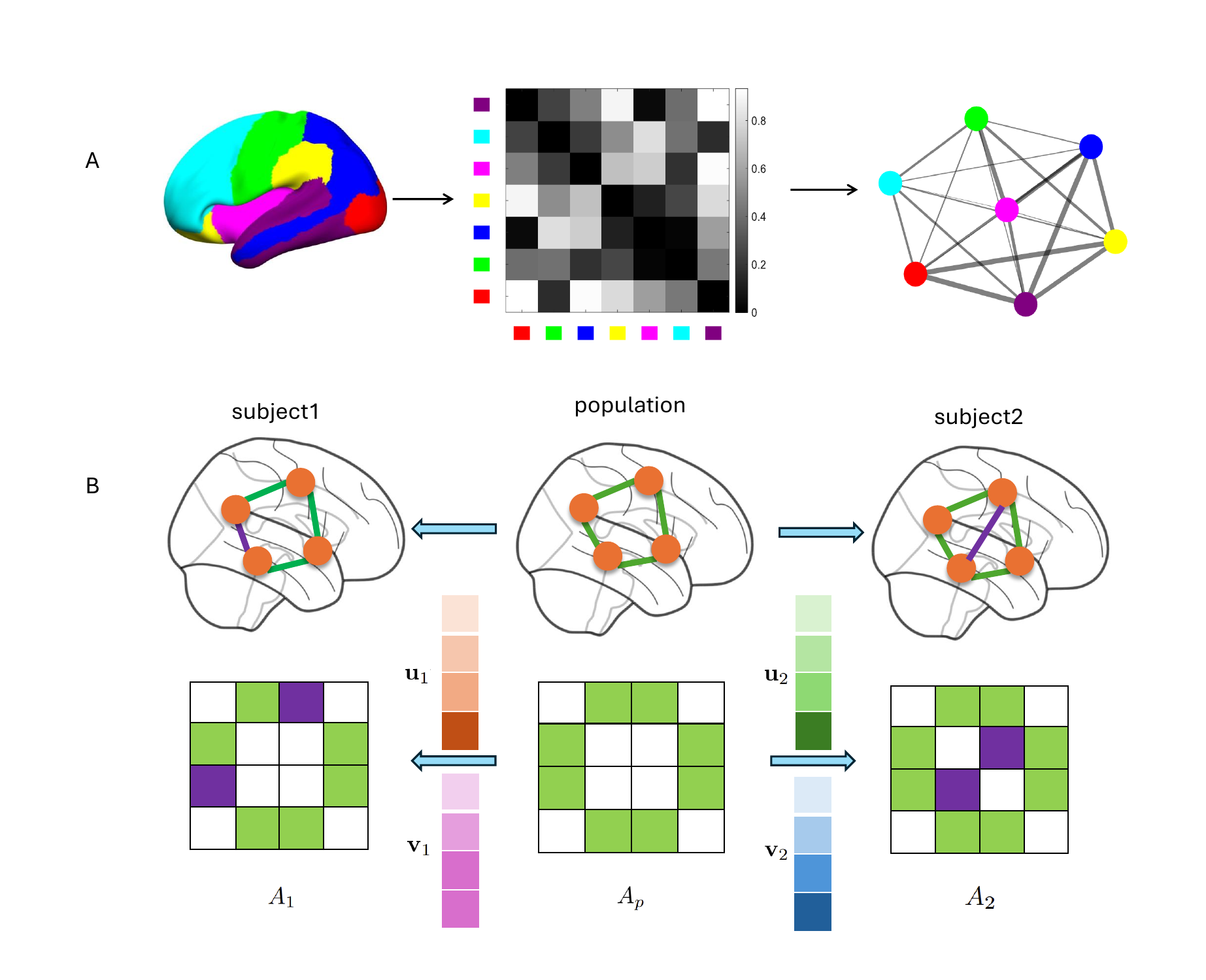}}
     \caption{A: Topological representation of the human brain as a graph $\mathcal{G}=(\mathcal{V}, \mathcal{E})$. Node set $\mathcal{V}$ comprises 7 functionally defined brain regions (Left). Edge set $\mathcal{E}$ is derived from a functional connectivity matrix, where connections between regions are thresholded for denoising (Middle). The undirected weighted graph $\mathcal{G}$ visualizes connectivity strengths, with edge widths proportional to the magnitude of entries in the functional connectivity matrix (Right). This graph structure enables spatially resolved modeling of Alzheimer's disease progression. B: A schematic representation of a simplified brain network as a graph, illustrating possible ways to derive a patient-specific adjacency matrix from the population-level adjacency matrix. The left arrow indicates that the topology remains unchanged, but the weights (FC values) on the edges differ, representing variations in the strength of functional connectivity between nodes. The right arrow illustrates a topology change, where a new edge is added to the graph, representing a functional connection strong enough to be included in the patient's network.}
    \label{fig:model-brain}
\end{figure}

We propose a spatiotemporal generalized Alzheimer's Disease Biomarker Cascade (ADBC) model that incorporates neuroimaging data and regional dynamics using graph theory and network science. This model can be used to capture the spatiotemporal progression of pathology in amyloid and taupathy, neuraldegenerion as well as cognitive decline. As illustrated in Fig.~\ref{fig:model-brain}A, we represent the brain's geometry as an undirected weighted graph $\mathcal{G}=(\mathcal{V}, \mathcal{E})$. Here, 68 nodes (vertices) correspond to DKT 68 atlas, while edges encode functional connectivity between regions. Edge weights, proportional to connectivity strength, are visualized as varying widths. This graph-based topology provides the foundation for our extended mathematical model, enabling spatially resolved modeling of disease progression while retaining the core dynamics of the ADBC framework. 

\rev{Now we are ready to extend the ADBC model to be a spatial discretized partial differential equation (PDE) of $A_\beta, \tau_\rho, N$ and the ordinal differential equation (ODE) of $C$ defined on the brain $\mathcal{G}$ as follows:
\begin{equation}\label{eq:ADBC_pde}
    \begin{aligned}
        &\frac{d A_\beta}{dt} + D_{A_\beta} L^G A_\beta = \lambda_{A_\beta}A_\beta(K_{A_\beta} - A_\beta), \\
        &\frac{d \tau_\rho}{dt} + D_\tau L^G \tau = \lambda_{\tau_\rho A_\beta}A_\beta + \lambda_{\tau_\rho}\tau_\rho(K_{\tau_\rho}-\tau_\rho),\\
        &\frac{d N}{dt} + D_N L^G N = \lambda_{N\tau_\rho}\tau_\rho + \lambda_N N(K_N- N),\\  
    \end{aligned}
\end{equation}
and  
\begin{equation}
    \frac{dC}{dt} = \lambda_{CN} \int_G N dv + \lambda_C C(K_C-C),
\end{equation}
where the integration in last equation is defined as:
\begin{equation}
    \int_G N(v) dv =\frac{\sum_{v\in \mathcal{V}}d_v N(v)}{\sum_{v\in \mathcal{V}} d_v},
\end{equation} 
where $d_v$ is the degree of vertices $v\in \mathcal{V}$. This term is a global burden summary of neurodegeneration. $\lambda_{CN}$ represents how strongly degeneration burden translates into cognitive decline rate.

In the model, $A_\beta$ represents amyloid pathology, $\tau_\rho$ represents amyloid-related tau pathology (measured by tau-PET), $N$ represents neuronal dysfunction/loss, and $C$ represents cognitive impairment. $D_{A_\beta}$, $D_\tau$, $D_N$ $\lambda_{CN}$, $\lambda_{C}$ and $K_C$ are 6 scalar parameters and  $\lambda_{A_\beta}$, $\lambda_{\tau_\rho}$, $\lambda_N$, $\lambda_{\tau_\rho A_\beta}$ $\lambda_{N\tau_\rho}$, $K_{A_\beta}$, $K_{\tau_\rho}$ and $K_N$ are 8 region-specific parameters which are graph functions $\mathcal{V}\rightarrow R$ defined on 68 vertices. $D_{A_\beta}$, $D_\tau$ and $D_N$ characterize the diffusion properties for $A_{\beta}$, $\tau$ and $N$, respectively. $\lambda_{A_\beta}$, $\lambda_{\tau_\rho}$, $\lambda_N$ and $\lambda_C$ reflect the logistic growth rates of the various biomarker cascades. $\lambda_{\tau_\rho A_\beta}$ and $\lambda_{N\tau_\rho}$ and $\lambda_{CN}$ reflects linear growth rates of the biomarkers and determine the influence of various factors on the time-of-onset of the subsequent biomarker cascades. $K_{A_\beta}, K_{\tau_\rho}, K_N$ and $K_C$, represent the biomarker carrying capacities respectively. We would like to estimate these model parameters using the longitudinal neuroimaging data. $L^G$ is the patient-specific graph Laplacian corresponding to the patient's specific brain functional connectivity network. Note that the graph Laplacian matrix is symmetric positive semi-definite which can be treated as the discretize version of $-\Delta$ using finite difference method with Neumann or periodic boundary condition. We will discuss how to learn patient specific graph Laplacian $L^G$ from limited functional connectivity data below. 

}

\subsection{Learning patient-specific functional connectivity matrix}
The graph Laplacian $L^G$ of a undirected weighted graph $\mathcal{G}=(\mathcal{V}, \mathcal{E})$ is defined as the difference between the corresponding degree matrix $D$ and adjacency matrix $A$ 
\begin{equation}\label{eq:graph-Laplacian}
    L^G=D-A,
\end{equation}
where $A=(a_{ij})\in R^{68 \times 68}$, the adjacency matrix which is symmetric, is collection of weights $a_{ij}$ assigned for connected node pairs $(n_i, n_j)\in \mathcal{E}$. 

The corresponding degree matrix $D$ that is, the number of edges attached to each node defined as follow:
\begin{equation}
    D_{ij} = \begin{cases}
        \sum_{k=1}^{68} a_{ik}, \quad &if \quad i=j,\\
        0 \quad &if \quad i\neq j.
    \end{cases}
\end{equation}
We use the functional connectivity (FC) matrix of 68 regions to compute the adjacency matrix $A$ of brain graph $\mathcal{G}$ after statistical truncation for denoising purpose\cite{ottoy2024tau}. More precisely, we treat the two subregions are connected where the corresponding Pearson correlation coefficient value in FC matrix exceed 0.75 and the corresponding $p$ value less than $1e-5$. Consequently, an edge is established between such pairs of nodes in the graph representation of $\mathcal{G}$. The corresponding adjacency matrix is weighted by the functional connectivity values after removing self-connections. 

FC data was derived by calculating Pearson's correlation coefficients between time series extracted from regions defined by the DKT 68 atlas\cite{DESIKAN2006968}, based on resting-state fMRI data. fMRI scans require expensive equipment and technical expertise. Conducting repeated scans over time (longitudinal studies) adds significantly to the cost. fMRI generates large datasets, espencially when acquired longitudinally. Processing these datasets to extract meaningful informations requires substantial computational resources.
The size of fMRI datasets necessitates significant storage capacity.
Patient data must be handled in compliance with strict privacy regulations, adding to the cost and complexity. Hence, a novel mathematical model/method for learning patient specific FC matrix (therefore \rev{graph} Laplacian) from limited FC data collected from public dataset is necessary.

We propose a qualitative method to learn a patient-specific adjacency matrix (FC matrix), defined as:
\begin{equation}\label{eq:adjacency-matrix}
    A = A_p + \frac{1}{2}(\bu\bv^T + \bv\bu^T) - \text{diag}\{\bu.\bv\},
\end{equation}
where \(\bu.\bv\) represents the elementwise product of \(\bu\) and \(\bv\), resulting in a vector. The term \(A_p = \frac{1}{K}\sum_{k=1}^K A^{(k)}\) denotes the population-level adjacency matrix derived from a limited population group. The rank-two matrix \(\frac{1}{2}(\bu\bv^T + \bv\bu^T)\) approximates the difference between a specific patient's connectivity and the population mean, while the diagonal term \(-\text{diag}\{\bu.\bv\}\) ensures self-connections are removed. The parameters \(\bu\) and \(\bv\), which are unit column vectors in \(\mathbb{R}^{68}\), quantify the patient-specific functional connectivity variation between different node pairs. \rev{The latent vectors $\bu$ and $\bv$ are introduced as a parsimonious parameterization of patient-specific deviations from the population-level connectivity network. They should not be interpreted as directly measurable biological quantities or clinical biomarkers. Instead, their role is analogous to latent factors in low-rank matrix factorization, providing a compact representation of individualized functional connectivity variations while maintaining model identifiability and computational tractability. In this framework, $\bu$ and $\bv$ generate subject-specific perturbations of the population adjacency matrix and consequently define personalized graph Laplacians for disease progression modeling.}

 If $\bu=\bv$, then, \eqref{eq:adjacency-matrix} degenerate to the case with rank one approximation as follows:
\begin{equation}\label{eq:rank-one}
    A = A_p + \bu\bu^T - \text{diag}\{\bu^2\}.
\end{equation}
We propose using a rank-two matrix instead of a rank-one matrix to improve expressive capability and to address limitations associated with rank-one representations. A rank-one matrix, such as \(\bu\bu^T\) with the constraint \( \|\bu\|_2 = 1 \), introduces unnecessary coupling between weights corresponding to different node pairs \((n_i, n_j)\) and \((n_k, n_l)\). For example:
\begin{equation}
   \bu\bu^T = \begin{bmatrix}
        u_1^2 & u_1u_2 & u_1u_3 & \cdots & u_1u_n \\
        u_2u_1 & u_2^2 & u_2u_3 & \cdots & u_2u_n \\
        \vdots & \vdots &  \vdots & \ddots & \vdots \\
        u_nu_1 & u_nu_2 & u_nu_3 & \cdots & u_n^2 \\
    \end{bmatrix},
\end{equation}
with the constraint \( \sum_{i=1}^{n} u_i^2 = 1 \). 

In contrast, the rank-two representation \(\bu\bv^T + \bv\bu^T\) provides greater flexibility:
\begin{equation}
\begin{bmatrix}
2u_1 v_1 & u_1 v_2 + v_1 u_2 & \cdots & u_1 v_n + v_1 u_n \\
u_2 v_1 + v_2 u_1 & 2u_2 v_2 & \cdots & u_2 v_n + v_2 u_n \\
\vdots & \vdots & \ddots & \vdots \\
u_n v_1 + v_n u_1 & u_n v_2 + v_n u_2 & \cdots & 2u_n v_n
\end{bmatrix},
\label{eq:symmetric_uv}
\end{equation}
\rev{where the unit constraints \( \|\bu\|_2 = \|\bv\|_2 = 1 \) are imposed to enhance numerical stability and prevent stiffness in solving ODE problems.}

This representation offers significantly enhanced expressive capability compared to \(\bu\bu^T\). For instance, in the rank-one matrix \(\bu\bu^T\), the weight for the connection between nodes \(n_1\) and \(n_2\) (\(u_1u_2\)) is tightly constrained by the weight for the connection between \(n_5\) and \(n_6\) (\(u_5u_6\)) due to the unit norm constraint \( \|\bu\|_2 = 1 \). In reality, the strength of the connection between \(n_1\) and \(n_2\) should not inherently depend on the strength of the connection between \(n_5\) and \(n_6\). The rank-two representation \(\bu\bv^T + \bv\bu^T\) overcomes this limitation, enabling a more accurate and flexible modeling of patient-specific FC.

As illustrated in Fig.~\ref{fig:model-brain}B, the brain network is represented as a graph with four nodes. The middle panel represents the population-level brain network, while individual deviations from this representation can be categorized into two fundamental cases:
\begin{enumerate}[label=(\alph*)]
    \item \textit{edge weight changes}: The topology remains unchanged, but the weights on the \rev{existing} edges differ from the population adjacency matrix. There exits an edge $e_i$ between nodes $n_k$ and $n_h$, and the weight is modified as $a_{kh}+\frac{1}{2}(u_kv_h+v_ku_h)-\delta_{kh}u_kv_h$ with constraint $-a_{kh}\leq \frac{1}{2}(u_kv_h+v_ku_h)-\delta_{kh}u_kv_h\leq 1-a_{kh}$. Note that the entries of $\bu, \bv$ can be negative values as long as they satisfy constraint so that the weight valued could be increased or decreased. 
    \item \textit{topology changes}: New edges are added, representing additional functional connectivity between regions. A new edge $e_i^{new}$ is lighted up between nodes $n_l$ and $n_m$ with the associated weight $a_{lm}=\frac{1}{2}(u_kv_h+v_ku_h)-\delta_{kh}u_kv_h$ with constraint $0\leq \frac{1}{2}(u_kv_h+v_ku_h)-\delta_{kh}u_kv_h\leq 1$.
\end{enumerate}
Luckily, the expression \eqref{eq:adjacency-matrix} can cover both cases. As shown in Fig.~\ref{fig:model-brain}B: (a) The left arrow indicates that the topology is identical to the population network, but the strength of functional connectivity (edge weights) varies indicated by a purple color of edge. (b) The right arrow illustrates a change in topology, where a new edge appears, representing functional connectivity between two regions strong enough to warrant inclusion. 
With this model, we address the challenges of data collection and computational expense, enabling efficient analysis of patient-specific brain networks. \rev{With the graph Laplacian definition \eqref{eq:graph-Laplacian} and the patient-specific adjacency matrix formulation \eqref{eq:adjacency-matrix}, the corresponding patient-specific graph Laplacian is defined as }
\begin{equation}\label{eq:Lp-specific}
    L^G = L^G_p + \frac{1}{2}\text{diag}(\mathbf{u} \cdot (\mathbf{v}^T \mathbf{1}) + \bv\cdot (\bu^T\mathbf{1})) -  \frac{1}{2}(\bu\bv^T + \bv\bu^T),
\end{equation}

\rev{where $L_p^G$ denotes the population-level graph Laplacian given by 
\begin{equation}
    L^G_p=\frac{1}{K}\sum_{k=1}^K (D^{(k)}-A^{(k)})-=\frac{1}{K}\sum_{k=1}^K D^{(k)}-A_p
\end{equation}
 
To ensure symmetry of the connectivity network, we replace each measured adjacency matrix by its symmetrized form
$A^{(k)}=\frac{A^{(k) T}+A^{(k)}}{2}$ which mitigates asymmetries introduced by the measurement noise. For computational tractability and parameter identifiability, we assume that the patient-specific graph Laplacian $L^G$ remains fixed throughout the observation period.}

\subsection{Parameters inference by novel hierarchical structured training strategy}\label{subsec:HS}
Let $\by=(A_{\beta}, \tau_{\rho}, N, C)^T$ represents the solution vector obtained by solving model \eqref{eq:ADBC_pde} and $\theta$ denotes the collection of all model parameters, and $\by_0$ be the initial values of this model. $\Tilde{\by}(t_i)$ is the clinical data of a specific subject for given age $t_i$ and $\by(t_i; \theta)$ is the solution of the generalized ADBC model. The model parameters for each patient can be inferred by solving the constrained optimization problem: 

\begin{equation}
  \begin{aligned}
    \min_{\theta, \by_0, \bu, \bv} \mathcal{L}(\theta, \by_0, \bu, \bv),\\
      \text{subject to:} \quad a_{kh} + \frac{1}{2}(u_k v_h + v_k u_h)-\delta_{kh}u_kv_h \geq 0, \\
      a_{kh} + \frac{1}{2}(u_k v_h + v_k u_h)-\delta_{kh}u_kv_h \leq 1,\\
     \forall k, h=1, ..., 68,
\end{aligned}
\end{equation}
where the objective function is defined as 
\begin{equation}
     \mathcal{L}(\theta, \by_0, \bu, \bv):= \sum_{i=1}^M  \frac{||\by(t_i; \theta)-\Tilde{\by}(t_i)||_2^2}{||\Tilde{\by}(t_i)||_2^2}  + w ||\by(100; \theta)-1||_2^2,
\end{equation}
where $\bu=(u_1, ...,u_{68}) \text{ and } \bv=(v_1, ..., v_{68})$ are two unit vectors to learn patient-specific graph Laplacian, $a_{kh}$ is the $(k, h)$ element of adjacency matrix $A_p$. $\by_0$ is imposed as an optimized variable to avoid the overfitting of the measurement noise in $\Tilde{\by}_0$. 

This optimization problem is to minimize the $L^2$ loss on given clinical patient data points and constraint on $\by(100; \theta)=1$ as the penalty term which is used to ensure the biomarkers and cognitive decline increasing with severity. This optimization problem was solved for each subject in the cohort using all available biomarker and cognitive decline time points $t_i, i=1,..., M$. 

\begin{figure}[!t]
    \centering
    \includegraphics[width=0.96\linewidth]{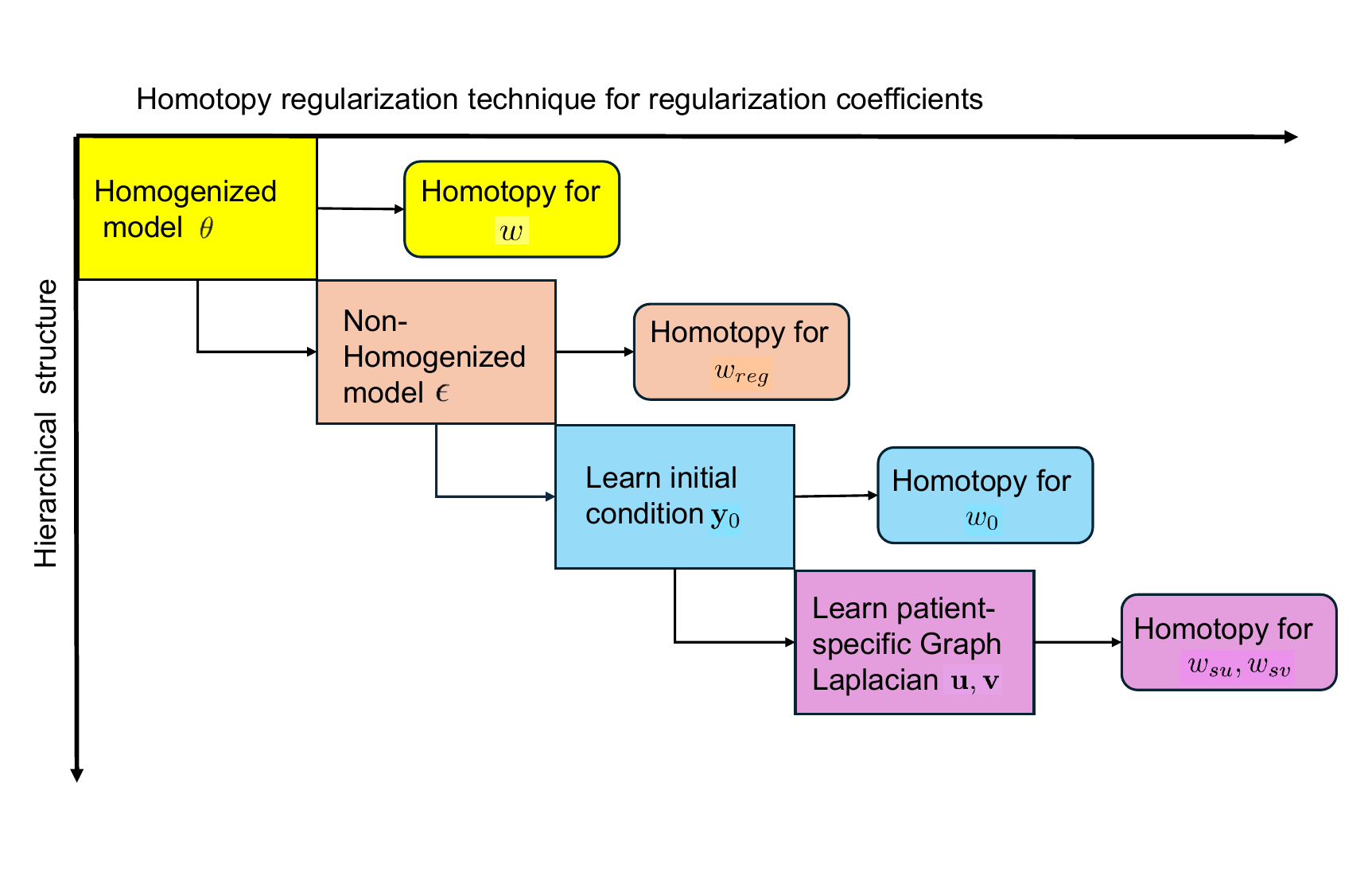}
    \caption{Diagram of the strategies for solving the complex original constrained optimization problem. The original optimization problem is splited into 4 optimization steps with a hierarchical structures as shown vertically. The homotopy reguraluraization technique is applied to every regularization coefficient $w$, $w_{reg}$, $w_0$, $w_{sv}$, and $w_{su}$ for every optimization step accordingly as shown horizontally.}
    \label{fig:HS}
\end{figure}

To address this challenging high-dimensional problem, we introduce a novel hierarchical structured training strategy incorporating a homotopy regularization technique to enhance the stability and efficiency of parameter learning as shown in Fig. \ref{fig:HS}. This hierarchical strategy consists of four sequential stages, where the solution obtained at each stage serves as the initial condition for the next, enabling a progressively refined learning of the extended ADBC model from neuroimaging data. \rev{By initializing each stage with the optimized solution from the previous stage, the framework progressively enlarges the parameter space while preserving previously learned information, thereby providing a monotonic refinement strategy in which additional model flexibility is introduced progressively, allowing the optimization to maintain or improve upon the fitting performance achieved at earlier stages.} To illustrate this procedure concretely, we take the A$\beta$ equation as an example and describe the hierarchical approach in detail below. The equations for other biomarkers can be solved in an equation-by-equation manner \cite{hao2018eqbyeq}.

\begin{enumerate}
    \item \textbf{Homogenized Model (benchmark):} To get a good initial guess for model parameters $\theta$, we homogenize the model by homogenizing the regional variability. Namely, let $(D^s_{A_{\beta}}, \lambda^s_{A_{\beta}}, K^s_{A_{\beta}})$ and $A^s_{\beta 0}$ be scalars. Then, the model parameters of this homogeneized model is $\theta_{A_\beta}^s=(\bI^{68}D^s_{A_{\beta}}, \bI^{68}\lambda^s_{A_{\beta}}, \bI^{68}K^s_{A_{\beta}})$. Because different patients have different onset times, in order to give the initial conditions, we assume that the initial time for all people is 50 years old, and the corresponding initial value $\bI^{68}A^s_{\beta0}$ is a parameter that needs to be inferred.
    The optimization problem is formulated as:
    \begin{equation} 
        \begin{aligned}
        \min_{\theta_{A_\beta}, A_{\beta0}} \mathcal{L}(\theta_{A_\beta}, A_{\beta0}) &= \sum_{i=1}^{M_{A_\beta}} \frac{||A_\beta(t_i; \theta_{A_\beta}) - \tilde{A_{\beta}}_i||_2^2}{||\tilde{A_{\beta}}_i||_2^2}\\
        &+ w ||A_\beta(100; \theta_{A_\beta}) - 1||_2^2,
        \end{aligned}
    \end{equation}
    where coefficient $w$ for the penalty term enforce a sigmoid-like solution. The parameters $\theta_{A_\beta}^s$ and $A^s_{\beta0}$ are optimized using MATLAB's \texttt{fmincon} iteratively with randomly generated initial guess with bounds $D_{A_{\beta}} \in [0, 2]$, $\lambda_{A_{\beta}} \in [0, 2]$, $K_{A_{\beta}} \in [1, 2]$, and $A^s_{\beta0}\in [0, 1]$. And the initial condition of this equation is $\bI^{68}A^s_{\beta 0}:=A_{\beta}(t_0=50)$ with $\bI^{68}=ones(68, 1)$. The spatial discreized PDE model is solved by \texttt{ode45}.  \\

    \item \textbf{Non-homogenized model:} Extend homogenized parameters $\theta_{A_\beta}^s$ to non-homogenized parameters, denoted as $\theta_{A_\beta}^{68D}$ to account for regional variability. Let the sparse deviations from the scalar parameters denoted as $\epsilon_{A_\beta}=\theta_{A_\beta}^{68D}-\theta_{A_\beta}^{s}$, the optimization problem is formulated as:
    \begin{equation} 
    \begin{aligned}
        \min_{\epsilon_{A_\beta}} \mathcal{L}(\epsilon_{A_\beta}) &= \sum_{i} \frac{||A_\beta(t_i; \epsilon_{A_\beta} + \theta_{A_\beta}^s) - A_{\beta}(t_i)||_2^2}{||A_{\beta}(t_i)||_2^2} \\
        & + w ||A_\beta(100; \epsilon_{A_\beta} + \theta_{A_\beta}^s) - 1||_2^2\\
        & + w_{reg}||\epsilon_{A_\beta}||_1,
    \end{aligned}
    \end{equation}
    where $\theta_{A_\beta}^s=(\bI^{68}D^s_{A_{\beta}}, \bI^{68}\lambda^s_{A_{\beta}}, \bI^{68}K^s_{A_{\beta}})$ and $A^s_{\beta 0}$ are the solution of homogenized model of step 1. To avoid overfitting issue, an $L^1$ regularization term is imposed for the model parameters. Regularization weight $w_{reg}$ is tuned to balance fit and sparsity so that to avoid potential overfitting. This optimization problem is solved using MATLAB's \texttt{fmincon} starting with $\theta_{A_\beta}^s$ as the initial of the model parameters. \\

    \item \textbf{Learnable Initial Conditions:} Extend homogenized initial condition $A^s_{\beta 0}$ to non-homogenized vector to account for regional variability, denoted as $A^{68D}_{\beta 0}$. To avoid overfitting issue, an $L^1$ regularization term is imposed for initial condition parameters. Let the sparse deviation from the scalar initial condition denoted as $\epsilon_{A_{\beta 0}}=A_{\beta 0}^{68D}-\bI^{68}A_{\beta 0}^s$. The optimization problem is formulated as:
    \begin{equation}
        \begin{aligned}
        \min_{\epsilon_{A_{\beta 0}}} \sum_{i} \mathcal{L}(\epsilon_{A_{\beta 0}})= & \frac{||A_\beta(t_i; \theta^{68D}_{A_\beta}) - A_{\beta}(t_i)||_2^2}{||A_{\beta}(t_i)||_2^2}\\
        &+ w ||A_\beta(100; \theta_{A_\beta}^{68D}) - 1||_2^2\\
        &+ w_0||\epsilon_{A_{\beta 0}}||_1,
        \end{aligned}
    \end{equation}
    where $\theta^{68D}_{A_\beta}$ is the solution obtained from the previous step and keep fixed during the optimization in this step. \\

    \item \textbf{Patient-Specific FC matrix:} For previous 3 steps, population FC matrix is keep fixed. Now, we learn patient-specific FC matrix from PET scan data by incorporating low-rank variation of $L_p$, modeled as \eqref{eq:Lp-specific}. The constrained optimization problem is formulated as:
    \begin{equation}
    \begin{aligned}
        \min_{\bu, \bv} \mathcal{L}(\bu, \bv) &= \sum_{i} \frac{\|A_\beta(t_i; \theta_{A_\beta}^{68D}) - A_{\beta}(t_i)\|_2^2}{\|A_{\beta}(t_i)\|_2^2} \\
        &+ w \|A_\beta(100; \theta_{A_\beta}^{68D}) - 1\|_2^2 \\
        &+ w_u (\|\bu\|_2 - 1)^2 + w_v (\|\bv\|_2 - 1)^2 \\
        &+ w_{su} \|\bu\|_1 + w_{sv} \|\bv\|_1, \\
        \text{subject to: } & a_{kh} + \frac{1}{2}(u_k v_h + v_k u_h)-\delta_{kh}u_kv_h \geq 0,\\
        & a_{kh} + \frac{1}{2}(u_k v_h + v_k u_h)-\delta_{kh}u_kv_h \leq 1,\\
        &\forall k, h=1, ..., 68,
    \end{aligned}
    \end{equation}
    where $a_{kh}$ is the $kh$-th element of FC matrix $A_p$, $\bu=(u_1, ...,u_{68}), \bv=(v_1, ..., v_{68})$ are imposed to ensure stability and interpret ability of the model and optimized by MATLAB \texttt{fmincon} function.
\end{enumerate}

\subsection{Homotopy regularization technique for optimization stability}\label{subsubsec:HR-train}
At each optimization step, a regularization term with a corresponding coefficient is introduced to prevent overfitting -- a critical requirement for ensuring the learned model's generalizability. However, directly setting the regularization coefficient to a small target value (often necessary for high model fidelity) leads to training instability due to the complex, non-convex loss landscape. To address this, we propose a homotopy regularization technique inspired by numerical homotopy continuation -- a method in computational mathematics that solves hard problems by gradually transforming them from simpler, related problems \cite{hao2018homotopy}.

The core idea involves solving a sequence of progressively harder optimization problems through a continuous deformation of the regularization coefficient. Specifically:
\begin{enumerate}
    \item Initialization: Begin with a large regularization coefficient (e.g., \(w_{reg}^0 = 1000\)), which simplifies the loss landscape by dominating the objective function, ensuring stable convergence.
    
    \item Progressive Refinement: For each stage, $w_{reg}^{k}, k=0, ..., n$ is fixed and after certain iteration steps of optimizations, one can reduce the  the regularization coefficient by a decay factor (e.g., \(w_{reg}^{k+1} = w_{reg}^k / 10\)). Namely, Gradually reduce the regularization coefficient by a decay factor at each stage.
    
    \item Warm-Start Propagation: Use the solution from the previous stage (larger \(\lambda\)) as the initial guess for the next stage (smaller \(\lambda\)), leveraging the continuity of the homotopy path.
\end{enumerate}

This approach effectively navigates the loss landscape by iteratively transitioning from a heavily regularized, convex-like regime to the target low-regularization regime, mitigating the risk of poor local minima or divergence.

\subsection{Training, testing and prediction evaluation protocol}
To evaluate our model's predictive capability on a per-patient basis, we employ a time-series split strategy using each patient's individual longitudinal data.

For a given patient with biomarker and cognitive decline measurements at $N$ time points ${t_1, t_2, ..., t_N}$, we implement the following protocol:

\textbf{Training (Parameter Inference)}: We use the first $N-1$ time points ${t_1, ..., t_{N-1}}$ as the training set. The model parameters are inferred by fitting the PDE dynamics to these historical data points.

\textbf{Testing (One-Step-Ahead Prediction)}: Using the parameters learned in the training step, we solve the PDE to simulate the biomarker dynamics and cognitive decline dynamics forward in time. We then evaluate the model's prediction at time $t_N$ (the last observed time point, which was held out during training). The testing accuracy is calculated by comparing this predicted value against the actual observed measurement at $t_N$. The accuracy  (\%) is defined as a normalized error-based similarity measure
\rev{\begin{equation}
	100 \times (1 - \frac{||\by-\hat{\by}||}{||\by||}), 
\end{equation}
where $\by\in R^{68}$ is the vector of quantities of interest and $\hat{\by}$ is the corresponding model prediction. }

\textbf{Prediction (Future Forecasting)}: Once the model's short-term accuracy is validated on the test point, we use the same inferred parameters to solve the PDE for future times. Specifically, we predict the biomarker value at $t_N + \delta t$ (e.g., $\delta t$ years into the future).

This protocol ensures that the model's predictive performance is evaluated on data not used during training, providing a robust assessment of its generalization capability for individual patients. 

\subsection{Sensitivity analysis of model parameters}
In this part, we introduce the sensitivity analysis of the model parameters which provides a better understanding of how the changes in the output of a model can be apportioned to different changes in the model parameters \cite{saltelli1995use}. Sobol method \cite{sobol2001global}, a variance-based sensitivity estimates for nonlinear mathematical models, is employed. It obtains the contribution of each parameter to the variance of the quantities of interest (i.e. the model output $C(100)$ in our case). The sensitivity index quantifies a parameter's influence on the model output—the larger the index, the greater the parameter's impact, and thus the higher its importance in the model.

Consider a model in the form $Y = f(X_1, ..., X_k)$ with $Y$ a scalar output of the model, $X_i$ is the $i-$th parameter and $\bX_{\sim i}$ denotes all parameters but $X_i$. The effect on the model output by varying $X_i$ alone, called the first-order index of $X_i$, is defined as the normalized variance given below 
\begin{equation}
    S^{(i)}_1 = \frac{Var_{X_i}{(E_{\bX_{\sim i}}(Y|X_i)})}{Var(Y)}.
\end{equation}
The effect on the model output by varying $X_i$ and $X_j$ simultaneously and removes the effect of their individual first-order indexes, called the second-order index, is defined as 
\begin{equation}
    S^{(ij)}_2 =  \frac{Var_{X_i,X_j}{(E_{\bX_{\sim ij}}(Y|X_i, X_j)})}{Var(Y)} - S^{(i)}_1 - S^{(j)}_1.
\end{equation}
The total index of parameter $X_i$ measures the contribution to the output variance of $X_i$ including all variance caused by its interactions, of any order, with any other model parameters and it is defined as 
\begin{equation}
    S^{(i)}_T = \frac{E_{\bX_{\sim i}} [Var_{X_i}(Y|\bX_{\sim i})]}{Var(Y)} = 1 - \frac{Var_{\bX_{\sim i}}[E_{X_i}(Y|\bX_{\sim i})]}{Var(Y)}.
\end{equation}
The total index $S^{(i)}_T$ measures the total effects, i.e. first and higher order effects (interactions) of parameter $X_i$.


\section*{Competing interests}
The authors declare that they have no competing interests.

\section*{Data availability}
All multimodal neuroimaging and clinical data used in this study were obtained from the Alzheimer’s Disease Neuroimaging Initiative (ADNI; \url{http://adni.loni.usc.edu/}) following approval of our data use application. Regional A$\beta$-PET and tau-PET standardized uptake value ratios (SUVRs) were derived from the “UC Berkeley - AV45 analysis [ADNI1, GO, 2, 3] (version: 2020-05-12)” and “UC Berkeley - AV1451 analysis [ADNI1, GO, 2, 3] (version: 2022-04-26)” datasets. Cortical thickness measures were obtained from the “UCSF Cross-Sectional FreeSurfer (6.0) [ADNI3]” and “UCSF Cross-Sectional FreeSurfer (5.1) [ADNI1, GO, 2]” releases. The Mini-Mental State Examination (MMSE) scores were obtained from “Mini-Mental State Examination (MMSE) [ADNI1,GO,2,3,4]”. Resting-state fMRI data were collected from 3T scanners following standardized ADNI acquisition protocols, and functional connectivity was computed between DKT-68 atlas regions.

\section*{Code availability}
The simulation study code of hierarchical structured training incorporating with homotopy regularization is openly available at \url{https://github.com/chunyanlimath/AD-DigitalTwin-Model}. The sensitivity analysis code is available at \url{http://salib.readthedocs.io/en/latest/}.

\section*{Acknowledgments}
C. L. and W. H. were supported by the National Institute of General Medical Sciences (NIGMS) through Grant No. R35GM146894. W. H. was also supported by the National Science Foundation (NSF) through Grant No. DMS-2533995 and by the Huck Chair in AI Mathematical Modeling from The Pennsylvania State University’s Huck Institutes of the Life Sciences.  Authors C. L. and W. H. received salary support from the NIGMS and NSF grants described above. The funders had no role in study design, data collection and analysis, decision to publish, or preparation of the manuscript.


\section*{Author contributions}

CL: Conceptualization, Methodology, Software, Formal analysis, Visualization, Writing – original draft.

YM: Investigation, Resources, Visualization, Writing – review \& editing.

XL: Conceptualization, Supervision, Funding acquisition, Writing – review \& editing.

WH: Conceptualization, Methodology, Supervision, Project administration, Funding acquisition, Writing – review \& editing.

All authors reviewed and approved the final manuscript.


\bibliography{ref_plos_checked}

@article{ottoy2024tau,
  title={Tau follows principal axes of functional and structural brain organization in Alzheimer’s disease},
  author={Ottoy, Julie and Kang, Min Su and Tan, Jazlynn Xiu Min and Boone, Lyndon and Vos de Wael, Reinder and Park, Bo-yong and Bezgin, Gleb and Lussier, Firoza Z and Pascoal, Tharick A and Rahmouni, Nesrine and others},
  journal={Nature communications},
  volume={15},
  number={1},
  pages={5031},
  year={2024},
  publisher={Nature Publishing Group UK London}
}

@article{shankar2008amyloid,
  title={Amyloid-$\beta$ protein dimers isolated directly from Alzheimer's brains impair synaptic plasticity and memory},
  author={Shankar, Ganesh M and Li, Shaomin and Mehta, Tapan H and Garcia-Munoz, Amaya and Shepardson, Nina E and Smith, Imelda and Brett, Francesca M and Farrell, Michael A and Rowan, Michael J and Lemere, Cynthia A and others},
  journal={Nature medicine},
  volume={14},
  number={8},
  pages={837--842},
  year={2008},
  publisher={Nature Publishing Group US New York}
}

@article{han2024global,
  title={Global brain activity and its coupling with cerebrospinal fluid flow is related to tau pathology},
  author={Han, Feng and Lee, JiaQie and Chen, Xi and Ziontz, Jacob and Ward, Tyler and Landau, Susan M and Baker, Suzanne L and Harrison, Theresa M and Jagust, William J and Alzheimer's Disease Neuroimaging Initiative},
  journal={Alzheimer's \& Dementia},
  volume={20},
  number={12},
  pages={8541--8555},
  year={2024},
  publisher={Wiley Online Library}
}

@article{vanDyck2023lecanemab,
  title={Lecanemab in early Alzheimer's disease},
  author={van Dyck, Christopher H and Swanson, Chad J and Aisen, Paul and Bateman, Randall J and Chen, Christopher and Gee, Michelle and Kanekiyo, Michio and Li, David and Reyderman, Larisa and Cohen, Sharon and others},
  journal={New England Journal of Medicine},
  volume={388},
  number={1},
  pages={9--21},
  year={2023},
  publisher={Massachusetts Medical Society}
}

@article{sims2023donanemab,
  title={Donanemab in early symptomatic Alzheimer disease: The TRAILBLAZER-ALZ 2 randomized clinical trial},
  author={Sims, John R and Zimmer, Jennifer A and Evans, Cynthia D and Lu, Ming and Ardayfio, Paul and Sparks, JonDavid and Wessels, Alette M and Shcherbinin, Sergey and Wang, Hong and Nemykina, Emon and others},
  journal={JAMA},
  volume={330},
  number={6},
  pages={512--527},
  year={2023},
  publisher={American Medical Association}
}

@article{cummings2023pipeline,
  title={Alzheimer's disease drug development pipeline: 2023},
  author={Cummings, Jeffrey and Zhou, Yadi and Lee, Gail and Zhong, Kate and Fonseca, Jorge and Cheng, Fei},
  journal={Alzheimer's \& Dementia: Translational Research \& Clinical Interventions},
  volume={9},
  number={2},
  pages={e12385},
  year={2023},
  publisher={Wiley Online Library}
}

@article{honig2023aria,
  title={ARIA in patients treated with lecanemab in the phase 3 Clarity AD study},
  author={Honig, Lawrence S and Barakos, Jerome and Dhadda, Shobha and Kanekiyo, Michio and Reyderman, Larisa and Irizarry, Michael and Kramer, Lynn D and Swanson, Chad J},
  journal={Alzheimer's \& Dementia},
  volume={19},
  pages={e076105},
  year={2023},
  note={Abstract from the Alzheimer's Association International Conference}
}

@article{selkoe2016amyloid,
  title={The amyloid hypothesis of Alzheimer's disease at 25 years},
  author={Selkoe, Dennis J and Hardy, John},
  journal={EMBO Molecular Medicine},
  volume={8},
  number={6},
  pages={595--608},
  year={2016},
  publisher={EMBO Press}
}

@article{tolar2020path,
  title={The path forward in Alzheimer's disease therapeutics: A new generation of amyloid-targeting agents},
  author={Tolar, Martin and Abushakra, Susan and Sabbagh, Marwan},
  journal={Journal of Alzheimer's Disease},
  volume={78},
  number={1},
  pages={1--8},
  year={2020},
  publisher={IOS Press}
}

@article{cottrell2025computational,
  title={Computational Drug Repurposing for {Alzheimer}'s Disease via Sheaf Theoretic Population-Scale Analysis of snRNA-seq Data},
  author={Cottrell, Sean and Yoon, Seungmin and Wei, Xiaoqi and Dickson, Alex and Wei, Guo-Wei},
  journal={arXiv preprint arXiv:2509.25417},
  year={2025}
}

@article{jack2017defining,
  title={Defining imaging biomarker cut points for brain aging and {Alzheimer}'s disease},
  author={Jack Jr, Clifford R and Wiste, Heather J and Weigand, Stephen D and Therneau, Terry M and Lowe, Val J and Knopman, David S and Gunter, Jeffrey L and Senjem, Matthew L and Jones, David T and Kantarci, Kejal and others},
  journal={Alzheimer's \& Dementia},
  volume={13},
  number={3},
  pages={205--216},
  year={2017},
  publisher={Elsevier}
}

@article{torok2025directionality,
  title={Directionality bias underpins divergent spatiotemporal progression of {Alzheimer}-related tauopathy in mouse models},
  author={Torok, Justin and Mezias, Christopher and Raj, Ashish},
  journal={Alzheimer's \& Dementia},
  volume={21},
  number={5},
  pages={e70092},
  year={2025},
  publisher={Wiley Online Library}
}

@article{raj2025understanding,
  title={Understanding the complex interplay between tau, amyloid and the network in the spatiotemporal progression of {Alzheimer}’s Disease},
  author={Raj, Ashish and Torok, Justin and Ranasinghe, Kamalini},
  journal={Progress in Neurobiology},
  pages={102750},
  year={2025},
  publisher={Elsevier}
}

@article{raj2024spectral,
  title={Spectral graph model for {fMRI}: A biophysical, connectivity-based generative model for the analysis of frequency-resolved resting-state {fMRI}},
  author={Raj, Ashish and Sipes, Benjamin S and Verma, Parul and Mathalon, Daniel H and Biswal, Bharat and Nagarajan, Srikantan},
  journal={Imaging Neuroscience},
  volume={2},
  pages={1--24},
  year={2024},
  publisher={MIT Press 255 Main Street, 9th Floor, Cambridge, Massachusetts 02142, USA~…}
}

@article{shahriyari2022pde,
  title={A PDE model of breast tumor progression in MMTV-PyMT mice},
  author={Mohammad Mirzaei, Navid and Tatarova, Zuzana and Hao, Wenrui and Changizi, Navid and Asadpoure, Alireza and Zervantonakis, Ioannis K and Hu, Yu and Chang, Young Hwan and Shahriyari, Leili},
  journal={Journal of Personalized Medicine},
  volume={12},
  number={5},
  pages={807},
  year={2022},
  publisher={MDPI}
}

@article{kirshtein2020colon,
  title={Data driven mathematical model of colon cancer progression},
  author={Kirshtein, Arkadz and Akbarinejad, Shaya and Hao, Wenrui and Le, Trang and Su, Sumeyye and Aronow, Rachel A and Shahriyari, Leili},
  journal={Journal of Clinical Medicine},
  volume={9},
  number={12},
  pages={3947},
  year={2020},
  publisher={MDPI}
}

@article{le2021osteosarcoma,
  title={Investigating optimal chemotherapy options for osteosarcoma patients through a mathematical model},
  author={Le, Trang and Su, Sumeyye and Shahriyari, Leili},
  journal={Cells},
  volume={10},
  number={8},
  pages={2009},
  year={2021},
  publisher={MDPI}
}

@article{cang2018representability,
  title={Representability of algebraic topology for biomolecules in machine learning based scoring and virtual screening},
  author={Cang, Zixuan and Mu, Lin and Wei, Guo-Wei},
  journal={PLoS computational biology},
  volume={14},
  number={1},
  pages={e1005929},
  year={2018},
  publisher={Public Library of Science San Francisco, CA USA}
}

@article{kostelich2025mathematical,
  title={Mathematical modeling for glioblastoma treatment: scenario generation and validation for clinical patient counseling},
  author={Kostelich, Eric J and Xu, Yuan and Calder{\'o}n-Valero, Carlos and Harris, Duane C and Alcantar-Garibay, Oscar and Gomez-Castro, Gerardo and On, Thomas J and Dortch, Richard D and Kuang, Yang and Preul, Mark C},
  journal={Frontiers in oncology},
  volume={15},
  pages={1647144},
  year={2025}
}

@article{tursynkozha2025go,
  title={Go-or-grow-or-die as a framework for the mathematical modeling of glioblastoma dynamics},
  author={Tursynkozha, Aisha and Harris, Duane C and Kuang, Yang and Kashkynbayev, Ardak},
  journal={Mathematical Biosciences},
  pages={109520},
  year={2025},
  publisher={Elsevier}
}

@article{baez2016mathematical,
  title={Mathematical models of androgen resistance in prostate cancer patients under intermittent androgen suppression therapy},
  author={Baez, Javier and Kuang, Yang},
  journal={Applied Sciences},
  volume={6},
  number={11},
  pages={352},
  year={2016},
  publisher={MDPI}
}

@article{huo2025oscillations,
  title={Oscillations in a tumor--immune system interaction model with immune response delay},
  author={Huo, Zhaoxuan and Huang, Jicai and Kuang, Yang and Ruan, Shigui and Zhang, Yuyue},
  journal={Mathematical Medicine and Biology: A Journal of the IMA},
  volume={42},
  number={2},
  pages={131--158},
  year={2025},
  publisher={Oxford Academic}
}

@article{dickerson2012mri,
  title={{MRI} cortical thickness biomarker predicts {AD}-like {CSF} and cognitive decline in normal adults},
  author={Dickerson, Bradford C and Wolk, David A},
  journal={Neurology},
  volume={78},
  number={2},
  pages={84--90},
  year={2012},
  publisher={Wolters Kluwer},
  doi={10.1212/WNL.0b013e31823efc6c},
  url={https://pubmed.ncbi.nlm.nih.gov/22189451/}
}

@article{dickerson2011alzheimer,
  title={Alzheimer-signature {MRI} biomarker predicts {AD} dementia: Regional cortical thinning in relation to future {AD} dementia},
  author={Dickerson, Bradford C and Bakkour, Akram and Salat, David H and Feczko, Eric and Pacheco, Jennifer and Greve, Douglas N and Grodstein, Francine and Wright, Christopher I and Blacker, Deborah and Rosas, H Diana and others},
  journal={Neurology},
  volume={76},
  number={2},
  pages={139--148},
  year={2011},
  publisher={Wolters Kluwer},
  doi={10.1212/WNL.0b013e3182166e96},
  url={https://www.neurology.org/doi/10.1212/WNL.0b013e3182166e96}
}

@article{harrison2021distinct,
  title={Distinct effects of beta-amyloid and tau on cortical thickness in cognitively healthy older adults},
  author={Harrison, Thomas M and Du, Rongxiang and Klencklen, G{\'e}raldine and Baker, Suzanne L and Jagust, William J},
  journal={Alzheimer's \& Dementia},
  volume={17},
  number={7},
  pages={1085--1096},
  year={2021},
  publisher={Wiley},
  doi={10.1002/alz.12249},
  url={https://alz-journals.onlinelibrary.wiley.com/doi/10.1002/alz.12249}
}

@article{keuss2024rates,
  title={Rates of cortical thinning in {Alzheimer}'s disease signature regions predict disease progression},
  author={Keuss, Stephanie E and Lane, Christopher A and Lungu, Olivia and Parker, Tim D and Coath, Will and Murray-Smith, Harriet and Poppe, Matthias and Hye, Abdullah and Frost, Chris and Schott, Jonathan M and others},
  journal={Journal of Neurology, Neurosurgery \& Psychiatry},
  volume={95},
  number={8},
  pages={748--755},
  year={2024},
  publisher={BMJ Publishing Group},
  doi={10.1136/jnnp-2023-333288},
  url={https://jnnp.bmj.com/content/95/8/748}
}

@article{racine2018personalized,
  title={The personalized {Alzheimer}’s disease cortical thickness index: characterization and replication in independent cohorts},
  author={Racine, Amber M and Clark, Lindsay R and Berman, Sarah E and Koscik, Rebecca L and Mueller, Kimberly D and Norton, Daniel and Nicholas, Carissa R and Blennow, Kaj and Zetterberg, Henrik and Carlsson, Cynthia M and others},
  journal={Alzheimer’s \& Dementia: Diagnosis, Assessment \& Disease Monitoring},
  volume={10},
  pages={400--410},
  year={2018},
  publisher={Elsevier},
  doi={10.1016/j.dadm.2018.02.003},
  url={https://www.sciencedirect.com/science/article/pii/S2352872918300150}
}

@article{mehta2024early,
  title={Early-onset {Alzheimer}’s disease {MRI} signature: a replication and extension},
  author={Mehta, Rajesh I and Wolf, Alice and Price, Laura and Kelly, Luke and Yousuf, Mohammed and van Westen, Danielle and Lindberg, Olof and Ferreira, Daniel and Hansson, Oskar and Westman, Eric},
  journal={Cerebral Cortex},
  volume={34},
  number={12},
  pages={bhae475},
  year={2024},
  publisher={Oxford University Press},
  doi={10.1093/cercor/bhae475},
  url={https://academic.oup.com/cercor/article/34/12/bhae475/7930075}
}

@article{hao2018homotopy,
  title={A homotopy method for parameter estimation of nonlinear differential equations with multiple optima},
  author={Hao, Wenrui},
  journal={Journal of Scientific Computing},
  volume={74},
  number={3},
  pages={1314--1324},
  year={2018},
  publisher={Springer}
}

@article{thal2002phases,
  title={Phases of {A$\beta$} deposition in the human brain and its relevance for the development of {AD}},
  author={Thal, Dietmar R and Rüb, Ursula and Orantes, M and Braak, Heiko},
  journal={Neurology},
  volume={58},
  number={12},
  pages={1791--1800},
  year={2002},
  doi={10.1212/WNL.58.12.1791}
}

@article{singh2006spatial,
  title={Spatial patterns of cortical thinning in mild cognitive impairment and {Alzheimer}'s disease},
  author={Singh, V and Chertkow, H and Lerch, JP and Evans, AC and Dorr, AE and Kabani, NJ},
  journal={Brain},
  volume={129},
  number={11},
  pages={2885--2893},
  year={2006},
  doi={10.1093/brain/awl256}
}

@article{du2007different,
  title={Different regional patterns of cortical thinning in {Alzheimer}'s disease and frontotemporal dementia},
  author={Du, AT and Schuff, N and Kramer, JH and et al.},
  journal={Brain},
  volume={130},
  number={4},
  pages={1159--1166},
  year={2007},
  doi={10.1093/brain/awm016}
}

@article{planche2022mri,
  title={Structural progression of {Alzheimer}’s disease over decades: the {MRI} staging scheme},
  author={Planche, V and Manj{\'o}n, JV and Mansencal, B and et al.},
  journal={Brain Communications},
  volume={4},
  number={3},
  pages={fcac109},
  year={2022},
  doi={10.1093/braincomms/fcac109}
}

@article{johnson2016tau,
  title={Tau {PET} imaging in aging and early {Alzheimer}’s disease},
  author={Johnson, KA and Schultz, A and Betensky, RA and et al.},
  journal={Annals of Neurology},
  volume={79},
  number={1},
  pages={110--119},
  year={2016},
  doi={10.1002/ana.24546}
}

@article{collij2022spatial,
  title={Spatial-temporal patterns of $\beta$-amyloid accumulation: a subtype and stage inference model analysis},
  author={Collij, Lyduine E and Salvad{\'o}, Gemma and Wottschel, Viktor and Mastenbroek, Sophie E and Schoenmakers, Pierre and Heeman, Fiona and Aksman, Leon and Wink, Alle Meije and Berckel, Bart NM and van de Flier, Wiesje M and others},
  journal={Neurology},
  volume={98},
  number={17},
  pages={e1692--e1703},
  year={2022},
  publisher={Lippincott Williams \& Wilkins Hagerstown, MD}
}

@article{ossenkoppele2016tau,
  title={Tau {PET} patterns mirror clinical and neuroanatomical variability in {Alzheimer}’s disease},
  author={Ossenkoppele, Rik and Schonhaut, David R and Sch{\"o}ll, Michael and et al.},
  journal={Brain},
  volume={139},
  number={5},
  pages={1551--1567},
  year={2016},
  doi={10.1093/brain/aww027}
}

@article{berron2021early,
  title={Early stages of tau pathology and its associations with functional connectivity, atrophy and memory},
  author={Berron, David and Vogel, Jacob W and Insel, Philip S and et al.},
  journal={Brain},
  volume={144},
  number={9},
  pages={2771--2783},
  year={2021},
  doi={10.1093/brain/awab114}
}

@article{scahill2002mapping,
  title={Mapping the evolution of regional atrophy in {Alzheimer}’s disease: unbiased analysis of serial {MRI}},
  author={Scahill, R I and Schott, J M and Stevens, J M and Rossor, M N and Fox, N C},
  journal={Proceedings of the National Academy of Sciences},
  volume={99},
  number={7},
  pages={4703--4707},
  year={2002},
  doi={10.1073/pnas.052587399}
}

@article{desikan2008mri,
  title={{MRI} measures of temporoparietal regions show differential rates of atrophy during prodromal {Alzheimer}’s disease},
  author={Desikan, R S and Fischl, B and Cabral, H J and et al.},
  journal={Neurology},
  volume={71},
  number={11},
  pages={819--825},
  year={2008},
  doi={10.1212/01.WNL.0000320055.57329.34}
}

@article{murray2015clinicopathologic,
  title={Clinicopathologic and \textsuperscript{11}C-PiB {PET} correlates of three {Alzheimer}’s disease subtypes: typical, limbic-predominant, and hippocampal-sparing},
  author={Murray, Melissa E and Graff-Radford, Neill R and Ross, Owen A and et al.},
  journal={Brain},
  volume={138},
  number={5},
  pages={1370--1381},
  year={2015},
  doi={10.1093/brain/awv050}
}

@article{korczyn2024alzheimer,
  title={Is {Alzheimer} disease a disease?},
  author={Korczyn, Amos D and Grinberg, Lea T},
  journal={Nature Reviews Neurology},
  volume={20},
  number={4},
  pages={245--251},
  year={2024},
  publisher={Nature Publishing Group UK London}
}

@article{hardy1992alzheimer,
  title={Alzheimer's disease: the amyloid cascade hypothesis},
  author={Hardy, John A and Higgins, Gerald A},
  journal={Science},
  volume={256},
  number={5054},
  pages={184--185},
  year={1992},
  publisher={American Association for the Advancement of Science}
}

@article{hao2025optimal,
  title={Optimal Control For Anti-Abeta Treatment in {Alzheimer}'s Disease using a Reaction-Diffusion Model},
  author={Hao, Wenrui and Kao, Chiu-Yen and Lee, Sun and Li, Zhiyuan},
  journal={arXiv preprint arXiv:2504.07913},
  year={2025}
}

@article{rabiei2025data,
  title={Data-Driven Modeling of Amyloid-beta Targeted Antibodies for {Alzheimer}'s Disease},
  author={Rabiei, Kobra and Petrella, Jeffrey R and Lenhart, Suzanne and Liu, Chun and Doraiswamy, P Murali and Hao, Wenrui},
  journal={arXiv preprint arXiv:2503.08938},
  year={2025}
}

@article{petrella2024personalized,
  title={Personalized Computational Causal Modeling of the {Alzheimer} Disease Biomarker Cascade},
  author={Petrella, Jeffrey R and Jiang, J and Sreeram, K and Dalziel, S and Doraiswamy, PM and Hao, W and Alzheimer's Disease Neuroimaging Initiative and others},
  journal={The journal of prevention of Alzheimer's disease},
  volume={11},
  number={2},
  pages={435--444},
  year={2024},
  publisher={Elsevier}
}

@article{zheng2022data,
  title={Data-driven causal model discovery and personalized prediction in {Alzheimer}'s disease},
  author={Zheng, Haoyang and Petrella, Jeffrey R and Doraiswamy, P Murali and Lin, Guang and Hao, Wenrui and Alzheimer’s Disease Neuroimaging Initiative},
  journal={NPJ digital medicine},
  volume={5},
  number={1},
  pages={137},
  year={2022},
  publisher={Nature Publishing Group UK London}
}

@article{hao2022optimal,
  title={Optimal anti-amyloid-beta therapy for {Alzheimer}’s disease via a personalized mathematical model},
  author={Hao, Wenrui and Lenhart, Suzanne and Petrella, Jeffrey R},
  journal={PLoS computational biology},
  volume={18},
  number={9},
  pages={e1010481},
  year={2022},
  publisher={Public Library of Science San Francisco, CA USA}
}

@article{vogel2020characterizing,
  title={Characterizing the spatiotemporal variability of {Alzheimer}’s disease pathology},
  author={Vogel, Jacob W and Young, Alexandra L and Oxtoby, Neil P and Smith, Ruben and Ossenkoppele, Rik and Strandberg, Olof T and La Joie, Renaud and Aksman, Leon M and J Grothe, Michel and Iturria-Medina, Yasser and others},
  journal={MedRxiv},
  pages={2020--08},
  year={2020},
  publisher={Cold Spring Harbor Laboratory Press}
}

@article{braak1991neuropathological,
  title={Neuropathological stageing of {Alzheimer}-related changes},
  author={Braak, Heiko and Braak, Eva},
  journal={Acta neuropathologica},
  volume={82},
  number={4},
  pages={239--259},
  year={1991},
  publisher={Springer}
}

@article{saltelli1995use,
  title={About the use of rank transformation in sensitivity analysis of model output},
  author={Saltelli, Andrea and Sobol, Ilya M},
  journal={Reliability Engineering \& System Safety},
  volume={50},
  number={3},
  pages={225--239},
  year={1995},
  publisher={Elsevier}
}

@article{sobol2001global,
  title={Global sensitivity indices for nonlinear mathematical models and their Monte Carlo estimates},
  author={Sobol, Ilya M},
  journal={Mathematics and computers in simulation},
  volume={55},
  number={1-3},
  pages={271--280},
  year={2001},
  publisher={Elsevier}
}

@article{DESIKAN2006968,
title = {An automated labeling system for subdividing the human cerebral cortex on {MRI} scans into gyral based regions of interest},
journal = {NeuroImage},
volume = {31},
number = {3},
pages = {968--980},
year = {2006},
issn = {1053-8119},
doi = {https://doi.org/10.1016/j.neuroimage.2006.01.021},
url = {https://www.sciencedirect.com/science/article/pii/S1053811906000437},
author = {Rahul S. Desikan and Florent Ségonne and Bruce Fischl and Brian T. Quinn and Bradford C. Dickerson and Deborah Blacker and Randy L. Buckner and Anders M. Dale and R. Paul Maguire and Bradley T. Hyman and Marilyn S. Albert and Ronald J. Killiany}
}

@article{scheltens2021alzheimer,
  title={Alzheimer's disease},
  author={Scheltens, Philip and De Strooper, Bart and Kivipelto, Miia and Holstege, Henne and Ch{\'e}telat, Gael and Teunissen, Charlotte E and Cummings, Jeffrey and van der Flier, Wiesje M},
  journal={The Lancet},
  volume={397},
  number={10284},
  pages={1577--1590},
  year={2021},
  publisher={Elsevier}
}

@article{nestor2006declarative,
  title={Declarative memory impairments in {Alzheimer}'s disease and semantic dementia},
  author={Nestor, Peter J and Fryer, Tim D and Hodges, John R},
  journal={Neuroimage},
  volume={30},
  number={3},
  pages={1010--1020},
  year={2006},
  publisher={Elsevier}
}

@article{hao2016mathematical,
  title={Mathematical model on {Alzheimer}’s disease},
  author={Hao, Wenrui and Friedman, Avner},
  journal={BMC systems biology},
  volume={10},
  number={1},
  pages={108},
  year={2016},
  publisher={Springer}
}

@article{petrella2019computational,
  title={Computational causal modeling of the dynamic biomarker cascade in {Alzheimer}’s disease},
  author={Petrella, Jeffrey R and Hao, Wenrui and Rao, Adithi and Doraiswamy, P Murali},
  journal={Computational and mathematical methods in medicine},
  volume={2019},
  number={1},
  pages={6216530},
  year={2019},
  publisher={Wiley Online Library}
}

@article{xu2025multiscale,
  title={A multiscale model to explain the spatiotemporal progression of amyloid beta and tau pathology in {Alzheimer}'s disease},
  author={Xu, Chunrui and Xu, Enze and Xiao, Yang and Yang, Defu and Wu, Guorong and Chen, Minghan},
  journal={International Journal of Biological Macromolecules},
  volume={310},
  pages={142887},
  year={2025},
  publisher={Elsevier}
}

@article{vosoughi2020mathematical,
  title={Mathematical models to shed light on amyloid-beta and tau protein dependent pathologies in {Alzheimer}’s disease},
  author={Vosoughi, Armin and Sadigh-Eteghad, Saeed and Ghorbani, Mohammad and Shahmorad, Sedaghat and Farhoudi, Mehdi and Rafi, Mohammad A and Omidi, Yadollah},
  journal={Neuroscience},
  volume={424},
  pages={45--57},
  year={2020},
  publisher={Elsevier}
}

@article{de2004mri,
  title={{MRI} and {CSF} studies in the early diagnosis of {Alzheimer}'s disease},
  author={De Leon, Mony J and DeSanti, S and Zinkowski, R and Mehta, PD and Pratico, D and Segal, S and Clark, C and Kerkman, D and DeBernardis, J and Li, J and others},
  journal={Journal of internal medicine},
  volume={256},
  number={3},
  pages={205--223},
  year={2004},
  publisher={Wiley Online Library}
}

@article{wiesman2021spatio,
  title={Spatio-spectral relationships between pathological neural dynamics and cognitive impairment along the {Alzheimer}'s disease spectrum},
  author={Wiesman, Alex I and Murman, Daniel L and May, Pamela E and Schantell, Mikki and Losh, Rebecca A and Johnson, Hallie J and Willet, Madelyn P and Eastman, Jacob A and Christopher-Hayes, Nicholas J and Knott, Nichole L and others},
  journal={Alzheimer's \& Dementia: Diagnosis, Assessment \& Disease Monitoring},
  volume={13},
  number={1},
  pages={e12200},
  year={2021},
  publisher={Wiley Online Library}
}

@article{patel2024mathematical,
  title={Mathematical modelling of {Alzheimer}’s disease biomarkers: Targeting Amyloid beta, Tau protein, Apolipoprotein E and Apoptotic pathways},
  author={Patel, Hetvi and Solanki, Nilay and Solanki, Arpita and Patel, Mehul and Patel, Swayamprakash and Shah, Umang},
  journal={American Journal of Translational Research},
  volume={16},
  number={7},
  pages={2777},
  year={2024}
}

@article{bertsch2021amyloid,
  title={The amyloid cascade hypothesis and {Alzheimer}’s disease: a mathematical model},
  author={Bertsch, Michiel and Franchi, Bruno and Meacci, Luca and Primicerio, Mario and Tesi, Maria Carla},
  journal={European Journal of Applied Mathematics},
  volume={32},
  number={5},
  pages={749--768},
  year={2021},
  publisher={Cambridge University Press}
}

@article{bossa2023multidimensional,
  title={A multidimensional ODE-based model of {Alzheimer}’s disease progression},
  author={Bossa, Mat{\'\i}as Nicol{\'a}s and Sahli, Hichem},
  journal={Scientific reports},
  volume={13},
  number={1},
  pages={3162},
  year={2023},
  publisher={Nature Publishing Group UK London}
}

@article{davodabadi2023mathematical,
  title={Mathematical model and artificial intelligence for diagnosis of {Alzheimer}’s disease},
  author={Davodabadi, Afsaneh and Daneshian, Behrooz and Saati, Saber and Razavyan, Shabnam},
  journal={The European Physical Journal Plus},
  volume={138},
  number={5},
  pages={474},
  year={2023},
  publisher={Springer}
}

@article{bertsch2023role,
  title={The role of {A$\beta$} and Tau proteins in {Alzheimer}’s disease: A mathematical model on graphs},
  author={Bertsch, Michiel and Franchi, Bruno and Tesi, Maria Carla and Tora, Veronica},
  journal={Journal of Mathematical Biology},
  volume={87},
  number={3},
  pages={49},
  year={2023},
  publisher={Springer}
}

@article{zhang2024discovering,
  title={Discovering a reaction--diffusion model for {Alzheimer}’s disease by combining PINNs with symbolic regression},
  author={Zhang, Zhen and Zou, Zongren and Kuhl, Ellen and Karniadakis, George Em},
  journal={Computer Methods in Applied Mechanics and Engineering},
  volume={419},
  pages={116647},
  year={2024},
  publisher={Elsevier}
}

@article{thompson2024alzheimer,
  title={Alzheimer’s disease and the mathematical mind},
  author={Thompson, Travis B and Vigil, Bradley Z and Young, Robert S},
  journal={Brain Multiphysics},
  volume={6},
  pages={100094},
  year={2024},
  publisher={Elsevier}
}

@article{wang2025learning,
  title={Learning Patient-Specific Spatial Biomarker Dynamics via Operator Learning for {Alzheimer}'s Disease Progression},
  author={Wang, Jindong and Mao, Yutong and Liu, Xiao and Hao, Wenrui},
  journal={arXiv preprint arXiv:2507.16148},
  year={2025}
}

@article{zhou2025accelerating,
  title={Accelerating Causal Network Discovery of {Alzheimer} Disease Biomarkers via Scientific Literature-based Retrieval Augmented Generation},
  author={Zhou, Xiaofan and Huang, Liangjie and Cheng, Pinyang and Yin, Wenpen and Zhang, Rui and Hao, Wenrui and Cheng, Lu},
  journal={arXiv preprint arXiv:2504.08768},
  year={2025}
}

@article{arezoumandan2022regional,
  title={Regional distribution and maturation of tau pathology among phenotypic variants of {Alzheimer}’s disease},
  author={Arezoumandan, Sanaz and Xie, Sharon X and Cousins, Katheryn AQ and Mechanic-Hamilton, Dawn J and Peterson, Claire S and Huang, Camille Y and Ohm, Daniel T and Ittyerah, Ranjit and McMillan, Corey T and Wolk, David A and others},
  journal={Acta neuropathologica},
  volume={144},
  number={6},
  pages={1103--1116},
  year={2022},
  publisher={Springer}
}

@inproceedings{sandell2025integrating,
  title={Integrating Event-Based and Network Diffusion Models to Predict Individual Tau Progression in Alzheimer's Disease},
  author={Sandell, Robin and Torok, Justin and Nagaragan, Srikantan and Ranasinghe, Kamalini G and Ma, Daren and Raj, Ashish},
  booktitle={Alzheimer's Association International Conference},
  year={2025},
  organization={ALZ}
}

@article{sandell2025back,
  title={Back to the Future: Predicting Individual Tau Progression in Alzheimer’s Disease},
  author={Sandell, Robin and Torok, Justin and Ranasinghe, Kamalini G and Nagarajan, Srikantan S and Raj, Ashish},
  journal={Research Square},
  pages={rs--3},
  year={2025}
}

@article{bougacha2025contributions,
  title={Contributions of connectional pathways to shaping Alzheimer’s disease pathologies},
  author={Bougacha, Salma and Roquet, Daniel and Landeau, Brigitte and Saul, Elise and Naveau, Mika{\"e}l and Sherif, Siya and Bejanin, Alexandre and Dhenain, Marc and Raj, Ashish and Vivien, Denis and others},
  journal={Brain Communications},
  volume={7},
  number={1},
  pages={fcae459},
  year={2025},
  publisher={Oxford University Press UK}
}

@article{tora2025network,
  title={A network-level transport model of tau progression in the Alzheimer’s brain},
  author={Tora, Veronica and Torok, Justin and Bertsch, Michiel and Raj, Ashish},
  journal={Mathematical Medicine and Biology: A Journal of the IMA},
  volume={42},
  number={2},
  pages={212--238},
  year={2025},
  publisher={Oxford University Press}
}

@article{butler2023choroid,
  title={Choroid plexus calcification correlates with cortical microglial activation in humans: a multimodal PET, CT, MRI study},
  author={Butler, Tracy and Wang, X Hugh and Chiang, Gloria C and Li, Yi and Zhou, Liangdong and Xi, Ke and Wickramasuriya, Nimmi and Tanzi, Emily and Spector, Edward and Ozsahin, Ilker and others},
  journal={American Journal of Neuroradiology},
  volume={44},
  number={7},
  pages={776--782},
  year={2023},
  publisher={American Journal of Neuroradiology}
}

@article{torok2023connectome,
  title={Connectome-based biophysics models of Alzheimer’s disease diagnosis and prognosis},
  author={Torok, Justin and Anand, Chaitali and Verma, Parul and Raj, Ashish},
  journal={Translational Research},
  volume={254},
  pages={13--23},
  year={2023},
  publisher={Elsevier}
}

@article{abdelnour2022advances,
  title={Advances in brain functional and structural networks modeling via graph theory},
  author={Abdelnour, Farras and Kuceyeski, Amy and Raj, Ashish and Iturria-Medina, Yasser and Deslauriers-Gauthier, Samuel},
  journal={Frontiers in Neuroscience},
  volume={16},
  pages={1031280},
  year={2022},
  publisher={Frontiers Media SA}
}

@article{sanami2025longitudinal,
  title={Longitudinal relationships among cerebrospinal fluid biomarkers, cerebral blood flow, and grey matter volume in individuals with a familial history of Alzheimer's disease},
  author={Sanami, Safa and Intzandt, Brittany and Huck, Julia and Villeneuve, Sylvia and Iturria-Medina, Yasser and Gauthier, Claudine J and Prevent-Ad Research Group and others},
  journal={Neurobiology of Aging},
  year={2025},
  publisher={Elsevier}
}

@article{hao2018eqbyeq,
  title={An equation-by-equation method for solving the multidimensional moment constrained maximum entropy problem},
  author={Hao, Wenrui and Harlim, John},
  journal={Communications in Applied Mathematics and Computational Science},
  volume={13},
  number={2},
  pages={189--214},
  year={2018},
  publisher={Mathematical Sciences Publishers}
}

\end{document}